\documentclass{jssc}

\usepackage{dsfont, enumitem, xcolor, url}
\usepackage[ruled,vlined,linesnumbered,algo2e]{algorithm2e}
\newif\ifrevisionmarks \revisionmarksfalse
\ifrevisionmarks
  
  \newcommand{\authornote}[1]{\par\noindent\fcolorbox{red}{red!5}{\parbox{0.96\linewidth}{\small\color{red!70!black}\textbf{Author decision needed:} #1}}\par}
\else
  
  \newcommand{\authornote}[1]{}
\fi
\theoremstyle{definition}
\newtheorem{assumption}[theorem]{Assumption}
\newcommand{\mc}[1]{\mathcal{#1}}
\newcommand{\ms}[1]{\mathsf{#1}}
\newcommand{\mr}[1]{\mathrm{#1}}

\newcommand{\mbb}[1]{\mathbb{#1}}
\newcommand{\mds}[1]{\mathds{#1}}
\newcommand{\rg}[1]{\mathring{#1}}
\newcommand{\mR}{\mathbb{R}}

\newcommand{\mN}{\mathbb{N}}
\newcommand{\xD}[1]{\mr{d} #1}
\newcommand{\xPP}[2]{\frac{\partial #1}{\partial #2}}
\newcommand{\bra}[1]{\left( #1 \right)}
\newcommand{\Bra}[1]{\left[ #1 \right]}
\newcommand{\BRA}[1]{\left\{ #1 \right\}}
\newcommand{\norm}[1]{\left\| #1 \right\|}
\newcommand{\ip}[2]{\langle #1, \, #2 \rangle}

\makeatletter
\def\@oddfoot{\hfill}
\newcount\shumeicount
\def\setshumei#1#2#3{%
  \shumeicount=\count0
  \def\@oddhead{%
    \raise-5pt\hbox to0pt{\vrule width\hsize height 0pt depth 0.4pt\hss}\relax
    \ifnum \shumeicount=\count0
      \raise-7pt\hbox to0pt{\vrule width\hsize height 0pt depth 0.4pt\hss}\relax
      #1
    \else
      \ifodd\count0
        #2
      \else
        #3
       \fi
     \fi
  }%
}
\makeatother
\makeatletter
\def\@oddfoot{\hfill}
\newcount\shujiaocount
\def\setshujiao{%
  \shujiaocount=\count0
  \def\@oddfoot{%
      \ifodd\count0
      \else
      \fi
  }%
}
\makeatother
\def\title#1#2#3#4{{
  \vspace*{0.3cm}
  \begin{flushleft} \Large\bf #1\end{flushleft}
  \vspace*{-0.2cm}
      \begin{flushleft}
      \bf #2
      \end{flushleft}
      \footnotetext{\hspace{-6mm} #3\\ #4}}}
\def\dshm#1#2#3#4
{\setshumei{J Syst Sci Complex (#1) #2:
{\thepage--\pageref{LastPage}}\hfill}
            {\hfill {\small #3}\hfill\hbox to0pt{\hss\thepage}}
            {\hbox to0pt{\thepage\hss}\hfill {\small #4}\hfill
            }
            \setshujiao}
\def\drd#1#2
{{\vskip 1cm\small \begin{flushleft}
 #1 \\
 #2\\
\copyright The Editorial Office of JSSC \&  Springer-Verlag GmbH
Germany 2021
\end{flushleft}}}
\def\hat{\widehat}
\def\tilde{\widetilde}
\def\bar{\overline}
\def\epsilon{\varepsilon}

\begin{document}

\title{Data-Driven Domain of Attraction Estimation via Convergent Koopman--Zubov Approximation}
{\uppercase{Tang} Wentao\\[4pt] {\normalfont\small November 22, 2026}}
{\uppercase{Tang} Wentao\\
\emph{Department of Chemical and Biomolecular Engineering, North Carolina State University, Raleigh, North Carolina, U.S.A.}
Email: \texttt{wtang23@ncsu.edu}.}
{This research was supported by the American Chemical Society Petroleum Research Fund under Grant No. 66911-DNI9 and the U.S. National Science Foundation under Grant 2414369.}

\pagestyle{fancy}\fancyhf{}\renewcommand{\headrulewidth}{0.4pt}
\fancyhead[C]{\small KOOPMAN--ZUBOV DOMAIN OF ATTRACTION}\fancyfoot[C]{\small\thepage}
\thispagestyle{plain}

\Abstract{This paper considers the problem of estimating the domain of attraction (DOA) of an asymptotically stable equilibrium, which is characterized by the support of a Zubov function. 
A Zubov--Koopman operator, i.e., a Koopman operator weighted by an accumulated cost, is defined and shown to be exponentially stable on a space of functions that vanish at the origin at least linearly (whenever the cost is sufficiently large away from the equilibrium). 
Consequently, the Zubov function is the unique invariant element on this space augmented with constants. 
Hence, by learning the Zubov--Koopman operator on a suitable reproducing kernel Hilbert space from snapshot data, an explicit kernel expression of the estimated Zubov function and an explicit DOA estimation algorithm are obtained. 
Such an estimation procedure is a highly efficient convex, closed-form regularized least-squares one, with a provable error bound proportional to the distance from the equilibrium and decreases with the fill distance of the sample. 
Numerical examples illustrate the accuracy and computational efficiency of the approach in comparison with existing neural network and dictionary-based methods. 
}


\section{Introduction}
For a nonlinear system, the study of a stable equilibrium point requires the characterization of its domain of attraction (DOA), namely the range of initial states such that the resulting orbits are ultimately attracted to the origin. 
The qualitative emergence and alternation of these rich invariant structures are studied in bifurcation theory \cite{khalil2002nonlinear}, and the design of a controller with maximal DOA volume is of interest in control theory \cite{chesi2011domain}.  
Conceptually, the certification of a maximal DOA is associated with the determination of a \emph{Zubov function} (or its transformed appearances). 
In Zubov's seminal monograph \cite{zubov1964method}, a state-dependent function that is $0$-valued at the origin, tends to $1$ as the state approaches the DOA boundary, and satisfies a Lyapunov-like descent property on the DOA, is conceived as the certificate for DOA. 
In \cite{vannelli1985maximal}, the problem is transformed to that of finding a so-called maximal Lyapunov function that tends to $\infty$ at the boundary. 
Computationally finding such a function, however, is difficult, if based only on the nonlinear governing equations, due to the lack of an analytical solution for the resulting Zubov's PDE or derived Hamilton--Jacobi PDE \cite{camilli2008control, meng2026characterization}. 
When a truncated series solution is used \cite{yu1967Nonlinear, dubljevic2002new}, only local validity can be aimed for and hence formal guarantee can be hardly obtained. The solution approaches based on partitioning the domain or simulating the system under a multitude of initial conditions tend be non-scalable to higher dimensions \cite{yuan2019estimation, kang2024data}. 

\par For low-order polynomial systems, the computation of the DOA can follow a sum-of-squares (SOS) programming approach \cite{topcu2009Local}, where the objective can be set to, e.g., a maximized DOA volume, and the variable to be sought is a nonnegative Lyapunov function whose sublevel set is forward-invariant. 
For non-polynomial systems, hierarchical relaxations are needed \cite{chesi2009estimating} to obtain the linear matrix inequality (LMI) constraints. 
In the spirit of certifying the stability with occupation measures \cite{vaidya2010nonlinear}, \cite{henrion2013convex} proposed an (infinite-dimensional) measure-theoretic optimization problem to obtain an initial measure with largest support set that transitions to a fixed final measure. 
When the system does not have a known model, or when such a model is non-polynomial or unavailable, \emph{data-driven approaches should be generally considered}. 

\par The use of neural networks as an surrogate of the Zubov's PDE solution was discussed in \cite{kang2024data}. Directly training a neural network as a maximal Lyapunov function whose sublevel set characterizes the DOA was only recently studied in \cite{harting2026locally}. Specifically, instead of directly training a neural ODE as the surrogate dynamics, the governing equation $f$ is supposed to be determined by three trainable neural networks according to a prefixed Zubov-type PDE. 
This idea was motivated by the use of neural networks with customized parameterization structures to impose control-theoretic properties, especially stability \cite{kolter2019learning, schlaginhaufen2021learning, kojima2022learning, wang2024monotone}. 
However, the neural network approach is intrinsically based on \emph{nonconvex learning}, which is not computationally light-weighted and the stochastic training result may not be consistent. Moreover, although universal approximation theorems can be established \cite{harting2026locally}, which mean that the error vanishes at extremely large neural networks, the finite-data error under deep learning is usually not well characterized. 
In \cite{liu2025physics}, physics-informed neural networks were used to solve Zubov PDEs and verified through satisfiability modulo theory (SMT) solvers; nevertheless, the computational cost is high. 
From both computational and theoretical viewpoints, it is desirable to have a \emph{convex learning approach} for DOA estimation. 

\par To this end, the concept of \emph{Zubov--Koopman operator} and its learning were proposed in \cite{meng2025learning}. 
Zubov--Koopman operator accounts for the nonlinear dynamics as well as a loss function associated with the deviation of the state from the origin. By putting the Zubov--Koopman operator on a constant-$1$ function ($\mds{1}$) recursively, the limit is a \emph{Zubov function} whose support estimates the (closure of) DOA. 
Since the Zubov--Koopman operator is a bounded linear operator on the space of continuous functions, it is learned via an extended dynamic mode decomposition (EDMD) routine with monomial dictionary functions, and this learning problem is convex. 
However, a major limitation of \cite{meng2025learning} was a lack of justification that the estimated Zubov--Koopman operator, when acting infinite times on $\mds{1}$, remains well-behaved, alike the true operator. 
Theoretically, such actions require that the Zubov--Koopman operator should \emph{have a spectral radius not exceeding $1$} and have a \emph{sole eigenvalue of $1$ on the unit circle}, with its associated eigenfunction being the Zubov function. 
This issue with the spectral structure of the Zubov--Koopman operator has not yet been resolved in the literature. 

\par Kernel methods offer another convex route. Giesl et al.\ approximated Lyapunov functions from noisy trajectory data by kernel regression \cite{giesl2020noisy}, and constructed certified Lyapunov functions from kernel approximations of Koopman eigenfunctions \cite{giesl2026kernel}. The present work differs in three respects: (i) the target is the Zubov function on the whole state space of interest, whose positivity set is the DOA, rather than a Lyapunov function on a sublevel set; (ii) only snapshot pairs are used without evaluations of $f$ or its derivatives; and (iii) the estimator is a closed-form fixed point of a learned linear operator, for which the error analysis reduces to operator perturbation and scattered-data approximation. 

\paragraph{Technical contributions.} This paper revisits the Zubov--Koopman operator of \cite{meng2025learning} and makes the following contributions.
\begin{enumerate}[label=(\roman*)]
    \item We show that, if the cost is sufficiently large away from a neighborhood of the equilibrium, the Zubov--Koopman semigroup is exponentially stable on the Banach space $C_{\mr{sect}}$ of continuous functions that vanish at least linearly at the origin (Lemma \ref{lem:Cw.stability}). Consequently, on $C_{\mr{sect}}$ augmented with constants, the Zubov function is the unique invariant element, and the iterates $Z_{\Delta t}^{\bar{k}}\mds{1}$ converge to it exponentially (Theorem \ref{th:invariant}). 
    \item We learn the operator on the RKHS associated with a linear--Sobolev kernel \cite{tang-ye2025koopman}, which the author in several previous works used for stability and dissipativity analysis \cite{tang-ye2026koopman, tang-ye2026dissipativity} and for systems with inputs \cite{tang2025koopman, morris-ye-tang2026}. 
    This RKHS embeds continuously into $C_{\mr{sect}}$. The learned operator equals $TZ_{\Delta t}$, where $T$ is a kernel ridge regression operator with $\|T\|\leq 1$, so all its powers are bounded uniformly in the sample size and the regularization. This yields an explicit kernel formula for the estimated Zubov function and an explicit DOA estimation procedure (Algorithm \ref{alg:DOA}). 
    \item We prove a \emph{sectorial} error bound, i.e., an error proportional to $|x|$, where the coefficient decreases with the fill distance of the sample (Theorem \ref{th:uniform}). The bound relies on a gradient sampling inequality (Lemma \ref{lem:sectorial}) and on the stability of the system. 
\end{enumerate}

\paragraph{Organization and notation.} Section \ref{sec:preliminaries} introduces the linear--Sobolev kernel, its associated RKHS, and the Zubov--Koopman operator. Section \ref{sec:operator} establishes the stability and invariance results on $C_{\mr{sect}}$. Section \ref{sec:learning} presents the learning procedure, the algorithm, and the error analysis. Numerical examples are given in Section \ref{sec:example} and conclusions in Section \ref{sec:conclusions}. All proofs are given in the Appendix, so that the paper is self-contained.

In the sequel, we always use lower-case Latin and Greek letters for scalars, vectors, scalar-valued functions, and vector-valued functions. Upper-case letters are used for matrices and operators. 
Blackboard fonts are used for $\mR^d$, $\mbb{C}$, $\mbb{N}$, or their subsets. 
Conventional function spaces use their typical upper-case notations, e.g., $C^s$: space of continuously differentiable functions, $L^2$: space of square-integrable functions, and $H^s$: Sobolev--Hilbert spaces. 
A RKHS with kernel $\kappa$ is denoted by $\mc{H}_\kappa$.  
Sans Serif lower-case letters stand for the data-driven representations of infinite-dimensional vectors (i.e., elements of a Hilbert space), while Sans Serif upper-case letters represent the data-driven representations of operators. $\mds{1}$ represents the function with a constant value of $1$.  
Norms are represented by $|\cdot|$ (on $\mR^d$ or $\mbb{C}$) or $\|\cdot\|$ on Hilbert spaces. Inner products are denoted as $\langle \cdot, \cdot\rangle$. Adjoint operators are denoted with an asterisk.

\section{Preliminaries}\label{sec:preliminaries}
In this section, useful definitions and conclusions from the existing RKHS theory, as well as the concept of Zubov--Koopman operator for DOA estimation, are introduced for later discussions. 
We are interested in a continuous-time system governed by ordinary differential equations:
\begin{equation}\label{eq:system}
    \dot{x} = f(x), \enspace f: \mbb{X}\to \mR^d. 
\end{equation}
The flow of the system passing through a time duration of $\Delta t$, namely the mapping that maps the initial condition $x(0)$ to the solution at the end-time $x(\Delta t)$, is denoted by $S_f^{\Delta t}$.
\begin{assumption}[State space]\label{assum:X}
	$\mbb{X}\subset\mR^d$ is a compact set with Lipschitz boundary and $0\in\operatorname{int}(\mbb{X})$; $f\in[C^1(\mbb{X})]^d$ with $f(0)=0$; $\mbb{X}$ is forward invariant, i.e., $S_f^t(x)\in\mbb{X}$ for all $x\in\mbb{X}$ and $t\geq 0$; and the origin is a locally exponentially stable equilibrium. 
\end{assumption}
The problem of interest is to estimate its DOA.
\begin{definition}[Domain of attraction]
	The DOA of \eqref{eq:system} is the subset of states that are ultimately attracted to the origin: $\mbb{A} = \BRA{ x\in \mbb{X}: \lim_{t\rightarrow \infty} S_f^t (x) =0}$. 
\end{definition}

\begin{remark}[On the forward invariance of $\mbb{X}$]\label{rem:taper}
	Forward invariance is needed so that the Koopman operator $g\mapsto g\circ S_f^t$ can be meaningfully defined as an operator on some function space over $\mbb{X}$. If the region $\mbb{X}$ is not forward invariant, one may replace $f$ by $\tilde f = \chi f$, where $\chi$ is a smooth cutoff factor that is equal to $1$ on a neighborhood $\mbb{X}_0\supset\mbb{X}$ and vanishing near $\partial\mbb{X}_1$ for some $\mbb{X}_1\supset \mbb{X}_0$. 
	Then $\mbb{X}_1$ becomes forward invariant for $\tilde f$, the trajectories of $\tilde f$ and $f$ coincide as long as they remain in $\mbb{X}_0$, and hence 
	$$\mbb{A}_{\tilde f}\cap\mbb{X} = \{x\in\mbb{X}: \text{the } f\text{-trajectory from } x \text{ converges to } 0 \text{ within } \mbb{X}_0\}, $$
	which coincides with the DOA of $f$ restricted to $\mbb{X}$ whenever $\mbb{X}_0$ contains the forward orbits of $\mbb{A}\cap\mbb{X}$. In the data-driven setting, this modification amounts to discarding the sample trajectories that leave $\mbb{X}_0$.
\end{remark}

\subsection{Sobolev--Hilbert Spaces and RKHS}
A \emph{Hilbert space} is an inner product space that is complete under its norm. Here we only consider real Hilbert spaces. 
An exemplar Hilbert space is the \emph{Sobolev--Hilbert space} $H^s(\mbb{X}):=W^{s,2}(\mbb{X})$. When $s\in \mN$, it is the space of such generalized functions $h: \mbb{X}\to\mR$ that have generalized derivatives up to $s$-th order, all of which being square-integrable over $\mbb{X}$:
\begin{equation*}
    \|h\|_{H^s}^2 := \sum_{\alpha\in \mN^d, \, |\alpha|\leq s} \int_{\mbb{X}} |\partial^\alpha h(x)|^2 \xD{x} < \infty.
\end{equation*}
When $s=0$, $H^0(\mbb{X}) = L^2(\mbb{X})$.
The Sobolev--Hilbert space $H^s(\mbb{X})$ contains functions that are sufficiently regular. This will be useful for defining the Koopman operator and Zubov--Koopman operator of dynamical systems. 

\begin{remark}
If $s>0$ is not an integer, $H^s(\mbb{X})$ is considered as a real interpolation space \cite{bergh2012interpolation} between $H^{\lfloor s\rfloor}(\mbb{X})$ and $H^{\lceil s\rceil}(\mbb{X})$:
\begin{equation*}
    H^s(\mbb{X}) = \Bra{ H^{\lfloor s\rfloor}(\mbb{X}), H^{\lceil s\rceil}(\mbb{X}) }_{\theta}, \enspace \theta=s-\lfloor s\rfloor \in (0, 1).  
\end{equation*}
Here the real interpolation between any two Hilbert spaces $\mc{H}_0$ and $\mc{H}_1$ can be defined as the space of $h\in \mc{H}_0+\mc{H}_1$ with 
\begin{equation*}
    \|h\|_{[\mc{H}_0, \mc{H}_1]_\theta}^2 := \int_0^\infty K(h, u)^2 u^{-(2\theta+1)} \xD{u}  <\infty,
\end{equation*}
in which the Peetre's $K$-functional is:
\begin{equation*}
        K(h, u) = \inf \BRA{ \|h_0\|_{\mc H_0} + u\|h_1\|_{\mc H_1}: h=h_0+h_1} .
\end{equation*}
\end{remark}

\par For unknown functions or operators, their learning from data can usually be conveniently conducted on \emph{reproducing kernel Hilbert spaces} (RKHSs) \cite{steinwart2008support}. 
We refer to a bivariate continuous function $\kappa:\mbb{X}\times\mbb{X}\rightarrow\mR$ as a \emph{Mercer kernel}, if the Gram matrix provided any finite number of points $\left\{ x_i \right\}_{i=1}^n$ in $\mbb{X}$, defined as $\ms \Phi_\kappa := \left[ \kappa\left(x_i, x_j\right) \right]_{i,j=1}^n$, is symmetric and positive semidefinite. 
Thus, for any two points $x$ and $x'$ in $\mbb{X}$, by letting $\ip{\kappa(x, \cdot)}{\kappa(x', \cdot)} = \kappa(x,x')$, an inner product is assigned on the span of $\{\kappa(x,\cdot): x\in \mbb{X}\}$ as a linear space of functions. 
Thus, the completion of this inner product space under its induced norm is a Hilbert space, which is called the RKHS with kernel $\kappa$:
\begin{equation*}
    \mathcal{H}_\kappa(\mbb{X}) = \overline{\operatorname{span}} \left\{ \kappa(x,\cdot) : x \in \mathbb{X} \right\}. 
\end{equation*}
We refer to the elements $\phi(x) := \kappa(x,\cdot)$, namely the kernel functions associated with each point $x\in \mbb{X}$, as the canonical feature of $x$ on the RKHS. 
An RKHS is guaranteed to have the \emph{reproducing kernel} property. That is, for any $g\in\mc{H}_\kappa(\mbb{X})$, we can write its evaluation at point $x$ as $g(x)=\ip{g}{\kappa(x,\cdot)} = \ip{g}{\phi(x)}$. 

\par To connect the Sobolev--Hilbert space to the concept of RKHS, we use the following conclusion \cite{wendland2004scattered}. 
\begin{definition}[Sobolev kernel]
    A Mercer kernel $\kappa$ on $\mbb{X}\subset \mR^d$ is said to be \emph{a radial kernel} if $\kappa(x,x')=\rho\left(\lvert x-x'\rvert\right)$ for a scalar-valued function $\rho:\mR_+\rightarrow\mR_+$. 
    If $\rho$ has a Fourier transform $\hat \rho$ such that
    \begin{equation*}
        c_1\left(1+\lvert \xi \rvert^2 \right)^{-s} \leq |\hat \rho (\xi)| \leq c_2 \left( 1+\lvert \xi \rvert^2 \right)^{-s}, \enspace \forall \xi \in \mR^d
    \end{equation*} 
    holds for some constants $c_2 \geq c_1 > 0$, then the kernel $\kappa$ is called a \emph{Sobolev kernel} (with smoothness index $s$) and denoted as $\kappa_{\mr{Sob}}$.  
\end{definition}

\begin{lemma}[Sobolev--Hilbert space as an RKHS]\label{lem:wendland}
    If $\mbb{X}$ has a Lipschitz boundary, then $H^s(\mbb{X}) \equiv \mc{H}_{\kappa_{\mr{Sob}}}
    (\mbb{X)}$, i.e., they contain the same elements and their norms (defined based on the squared integrals of derivatives and the RKHS calculus respectively) are equivalent.
\end{lemma}

\begin{remark}[Wendland kernel]
\label{rem:Wendland.kernel}
    It can be proven that if $\hat \rho(\xi) \propto (1+|\xi|^2)^{-s}$ is required, then the radial function $\rho$ is the Mat{\'{e}}rn covariance function:
    \begin{equation*}
        \rho(r) \propto \bra{\sqrt{2\nu}r}^\nu \mr{K}_\nu(\sqrt{2\nu}r), \enspace \nu = s-d/2, 
    \end{equation*}
    where $\mr K_\nu$ is the modified Bessel function of the second type. The distance $r = |x-x'|$ can be scaled by a constant ``bandwidth parameter'' $\sigma>0$. 
    An alternative construction, given in \cite{wendland2004scattered} and denoted by $\kappa^{\ms{W}}_{d, k}$, is constructed as follows. 
    First, for $l\in \mN$, let $\rho_l: \mR_+\to\mR_+$, $\rho\mapsto \max\{1-r, 0\}^l$. Using an operator $I$ on the space of polynomials supported within $[0, 1]$: $(Ig)(r) = \int_r^\infty r'g(r')\xD{r'}$, one can denote $\rho_{d,k} = I^k\rho_{\lfloor n/2 \rfloor + k+1}$ and let $\kappa_{d,k}^{\ms{Wen}}(x, x') = \rho_{d,k}(|x-x'|)$. 
    The definition can be posed for $d\geq 1$ if $k\in \mN$ and $d\geq 3$ if $k=0$.
    It is guaranteed that 
    \begin{equation*}
        \mc{H}_{\kappa_{d, k}^{\ms{Wen}}}(\mbb{X}) \equiv H^{(d+1)/2+k}(\mbb{X}), \text{ i.e., } s=k+1/2.
    \end{equation*} 
\end{remark}

\subsection{Linear--Sobolev Space and Linear--Sobolev Kernel}
\par In the literature \cite{kohne2025error, bold2025kernel}, the Sobolev-type RKHS $H^s(\mbb{X}) \equiv \mc{H}_{\kappa_{\mr{Sob}}}(\mbb{X})$ has been used. The idea is that, if the dynamical system in scope possesses sufficient smoothness, then the ``evolution'' of functions along the dynamics should be nicely restricted within the Sobolev space. 
However, in dynamical systems, a common concern is the properties of an \emph{equilibrium point} (which is also the case with this study on DOA estimation). A radial Sobolev kernel, however, is indifferent to such a special structure. 
Intuitively, one can easily note that $\kappa_{\mr{Sob}}(x,x') = \rho(|x-x'|)$ depends only on the distance between $x$ and $x'$, regardless whether either of them is close to the equilibrium. 
To the end of capturing both ``global property'' of the dynamical system, namely the smoothness, and the ``local property'' of an equilibrium, \cite{tang-ye2025koopman} proposed the following linear--Sobolev space as the RKHS with a linear--Sobolev kernel. 
The idea is reviewed below. 

\par A simple Mercer kernel on $\mbb{X}$ is the linear kernel. For systems with linear dynamics, it suffices to consider the RKHS with the linear kernel. Intuitively, it is suitable for capturing the nonlinear dynamics ``locally'' \cite{mezic2020spectrum}. 
\begin{definition}[Linear kernel]
    The linear kernel refers to $\kappa_\mr{lin}(x,x') = x^\top x'$. If $\mbb{X}\subset \mR^d$ contains $0$ in its interior, the resulting RKHS is the space of linear functions
    \begin{equation*}
        \mc{H}_{\kappa_\mr{lin}}(\mbb{X}) = \BRA{ x\mapsto c^\top x: c\in \mR^d} = \operatorname{span}\BRA{e_k}_{k=1}^d, 
    \end{equation*}
    where $e_k: x\mapsto x_k$ are the component projections. 
\end{definition}
By denoting $\ms{e}_k$ as the unit vector in $\mR^d$ that aligns with the $k$-th positive semi-axis, it can be clearly seen that $e_k = \kappa_{\mr{lin}}(\ms e_k, \cdot)$, and thus $\ip{e_k}{e_{k'}} = \kappa_{\mr{lin}}(\ms e_k, \ms e_{k'}) = \delta_{kk'}$ (Kronecker delta). 
Therefore, $\mc{H}_{\kappa_\mr{lin}}(\mbb{X})$ has all Euclidean projection mappings $\{e_k\}_{k=1}^d$ as its orthonormal basis. 
With both the linear kernel and the radial Sobolev kernel, we can multiply them to generate a new kernel. The validity of a product kernel is well-known in the RKHS theory \cite{steinwart2008support}. 
\begin{lemma}[Product kernel]
    Let $\kappa_1$ and $\kappa_2$ be two Mercer kernels on $\mbb{X}\subset \mR^d$. Then $\kappa=\kappa_1\kappa_2: (x,x')\mapsto \kappa_1(x,x') \kappa_2(x,x')$ is a Mercer kernel. 
    The RKHS with the product kernel is spanned by the products of elements in the constituent RKHSs:
    \begin{equation*}
        \mc{H}_\kappa(\mbb{X}) = \overline{\operatorname{span}}\BRA{ g_1g_2: g_1\in \mc{H}_{\kappa_1}(\mbb{X}), g_2\in \mc{H}_{\kappa_2}(\mbb{X}) }.
    \end{equation*}
    More precisely, $\mc{H}_\kappa(\mbb{X})$ is the image of the Hilbert tensor product $\mc{H}_{\kappa_1}(\mbb{X})\otimes\mc{H}_{\kappa_2}(\mbb{X})$ under the pointwise-product map $P: g_1\otimes g_2\mapsto g_1g_2$, with the quotient norm $\|g\|_{\mc{H}_\kappa} = \inf\{\|u\|_{\otimes}: Pu = g\}$. The identity $\ip{g_1\otimes g_2}{g_1'\otimes g_2'}_{\otimes} = \ip{g_1}{g_1'}_{\mc{H}_{\kappa_1}}\ip{g_2}{g_2'}_{\mc{H}_{\kappa_2}}$ holds on the tensor product.
\end{lemma}

\begin{definition}[Linear--Sobolev kernel]\label{def:linear--radial}
    Denote
    \begin{equation*}
        \rg\kappa(x,x') = \kappa_\mr{lin}(x,x')\kappa_{\mr{Sob}}(x,x') = \left( x^\top x' \right) \rho\left(\lvert x - x'\rvert\right),  
    \end{equation*}
    which is indeed a Mercer kernel on $\mbb X$. The canonical feature of any $x\in \mbb{X}$ on the $\mc H_{\rg\kappa}(\mbb X)$ is denoted as $\rg\phi(x) = \rg\kappa(x, \cdot)$. 
\end{definition}
\begin{corollary}[Linear--Sobolev space]
\label{cor:linear.Sobolev}
    Suppose that $\mbb{X}\subset \mR^d$ is a domain with Lipschitz boundary and contains the origin in its interior. 
    The following space, referred to as the \emph{linear--Sobolev space}:
    \begin{equation*}
        \rg H^s(\mbb{X}) = \left\{ \sum_{k=1}^d e_kg_k : g_k \in H^s(\mbb{X}) \right\}, \enspace \norm{ g }_{\rg H^s}^2 = \inf\BRA{\sum_{k=1}^d \norm{g_k}_{H^s}^2 : g = \sum_{k=1}^d e_kg_k,\ g_k\in H^s(\mbb{X})},
    \end{equation*} 
    is a Hilbert space and is equivalent to the RKHS with kernel $\rg\kappa$, i.e., $\rg H^s(\mbb{X}) \equiv \mc H_{\rg\kappa}(\mbb{X})$. 
\end{corollary}

\par Informally speaking, any member of the linear--Sobolev space is the linear combination of (at most $d$) products of a linear function and a Sobolev-type function. 
When $s>d/2$, by Sobolev embedding theorem, the Sobolev--Hilbert function space can be continuously embedded into a continuous function space. Thus, the linear--Sobolev space contains functions that are \emph{locally at least linear} near the origin, while still globally regular in the Sobolev sense. 
Moreover, it can be seen that 
\begin{enumerate}[label=(\roman*)]
    \item since $\rg\phi(0) = \rg\kappa(0, \cdot) = 0^\top (\cdot)\rho(|\cdot|)$ is the constant-$0$ function, we have $\|\rg\phi(0)\| =0$ on the RKHS; and more generally for all $x\in \mbb{X}$, 
    \item since $\|\rg\phi_x\|^2 = \rg\kappa(x,x)=|x|^2\rho(0)$, we have $\|\rg\phi(x)\|\propto |x|$. 
\end{enumerate}
In contrast, the Sobolev kernel would make $\|\phi_{\mr{Sob}}(x)\| = \kappa_{\mr{Sob}}(x, x) = \rho(0)$, which is a constant. In other hands, with a radial kernel, all points $x\in \mbb{X}$ are lifted onto the same sphere, and the origin does not correspond to the origin on the RKHS. 
With these properties, we say that the linear--radial kernel is \emph{origin-preserving}.

\subsection{Koopman Operator and Koopman Semigroup}
For the nonlinear system \eqref{eq:system}, Koopman operator provides a fully linearized description of the dynamics. 
\begin{definition}[Koopman operator and semigroup]
    Given any $\Delta t\geq0$, the \emph{Koopman operator} refers to the composition mapping:
    \begin{equation*}
        K_{\Delta t}: \mc{G}\to \mc{G}, \enspace g\mapsto g\circ S_f^{\Delta t}
    \end{equation*}
    defined on a function space $\mc{G}$ of state-dependent functions, supposing that $\mc{G}$ is indeed invariant under the composition. 
    The family of Koopman operators $\{K_{\Delta t}\}_{\Delta t \geq 0}$ forms a semigroup, called the \emph{Koopman semigroup}, as it clearly satisfies: (i) $K_0 = I$, and (ii) $K_{\Delta t_2}K_{\Delta t_1} = K_{\Delta t_1+\Delta t_2}$. 
\end{definition}

\begin{remark}[Adjoint operator]
\label{rem:adjoint}
    The adjoint operator of $T:\mc{G}\to \mc{G}$ is the operator $T^*: \mc{G}'\to\mc{G}'$ (where $\mc{G}'$ is the dual space of $\mc{G}$, i.e., the space of linear functionals from $\mc{G}$ to $\mR$) that satisfies $(T^*\ell)(g) = \ell(Tg), \enspace \forall g\in \mc{G}, \, \ell\in \mc{G}'$. 
    In the case where $\mc{G}$ is a Hilbert space, $\mc{G}' \equiv \mc{G}$ according to the Riesz--Fr{\'{e}}chet theorem. Thus, $\ip{T^*h}{g} = \ip{h}{Tg}$ for all $g, h\in \mc{G}$. 
    Suppose that the function space $\mc{G}$ is further a RKHS with kernel $\kappa$, i.e., $\mc{G} = \mc{H}_{\kappa}(\mbb{X})$, then it can be easily verified that the adjoints of the Koopman operators (also known as the \emph{Perron--Frobenius operators}) satisfy:
    \begin{equation*}
        K_{\Delta t}^* \phi(x) = \phi\bra{S_f^{\Delta t}(x)}, \enspace \forall x\in \mbb{X}, \, \Delta t \geq 0.
    \end{equation*}
\end{remark}

It should be justified that the Koopman semigroup is a strongly continuous semigroup. To this end, the conclusion on the boundedness of Koopman operator (for discrete-time systems) on the Sobolev--Hilbert spaces can be extended to continuous-time systems. 
The condition is that the dynamics should be $s$-smooth, i.e., $f$ componentwise belongs to $C^s(\mbb{X})$ \cite{kohne2025error}. 
On the linear--Sobolev space $\rg H^s(\mbb{X})$, to arrive at an analogous conclusion, the condition is that the dynamics, componentwise, should belong to the following class as a Banach space:
\begin{equation*}
\begin{aligned}
    \rg C^s(\mbb{X}) &= \BRA{p= e_1p_1+\cdots+e_dp_d: p_1, \cdots, p_d\in C^s(\mbb{X})}, \\
    \norm{p}_{\rg C^s} &= \inf\BRA{\max_{k=1,\cdots,d} \|p_k\|_{C^s} : p = \textstyle\sum_{k=1}^d e_kp_k}, \quad \|p_k\|_{C^s} = \max_{|\alpha|\leq s} \sup_{x\in \mbb{X}} \left\vert \partial^\alpha p_k(x) \right\vert. 
\end{aligned}
\end{equation*}
As for $\rg H^s$, the infimum over representations is needed for the norm to be well defined. Note that $|p(x)|\leq\sqrt{d}\,\|p\|_{\rg C^0}|x|$ by the Cauchy--Schwarz inequality, so $\rg C^0(\mbb{X})$ embeds in the space $C_{\mr{sect}}(\mbb{X})$ introduced in Section \ref{sec:operator}; the converse inclusion is false in general, and the two norms are not equal.

\begin{remark}
    It is not hard to see that if $p\in \rg C^s(\mbb{X})$ and $\mbb{X}$ is bounded, then $p(0)=0$ and $p\in C^s(\mbb{X})$. If $p(0)=0$, $p\in C^{s+1}(\mbb{X})$, and that $\mbb{X}$ is star-shaped (i.e., for any $x\in \mbb{X}$, the segment $[0, x]=\{\alpha x: 0\leq \alpha \leq 1\}$ is contained in $\mbb{X}$), then $p\in \rg C^s(\mbb{X})$. This is because $p(x) = \int_0^1 \nabla p(\alpha x)^\top x \xD{\alpha} = \sum_{k=1}^d x_k \int_0^1 \partial_{x_k} p(\alpha x) \xD{\alpha}$, where $\partial_{x_j} p(\alpha x)$, and hence the integral under the summation, is a $C^s$-class function of $x$. 
\end{remark}

\begin{lemma}[Strong continuity of Koopman semigroup]
    If $f\in \Bra{C^s(\mbb{X})}^d$, then $\{K_{\Delta t}\}_{\Delta t\geq 0}$ is a strongly continuous semigroup on $H^s(\mbb{X})$. If $f\in \Bra{\rg C^s(\mbb{X})}^d$, then $\{K_{\Delta t}\}_{\Delta t\geq 0}$ is a strongly continuous semigroup on $\rg H^s(\mbb{X})$.
\end{lemma}
The above lemma was proven in \cite{ye2026}. Now that $\rg H^s(\mbb{X})$ is an RKHS with linear--radial kernel $\rg\kappa$, we say that the Koopman semigroup is a strongly continuous semigroup on $\mc{H}_{\rg\kappa}(\mbb{X})$. As such, the adjoint relation in Remark \ref{rem:adjoint} holds.

\subsection{Zubov--Koopman Operator and Semigroup}
To study the DOA, it is necessary to use a specific function to indicate the asymptotic behavior, i.e., the convergence to the origin. We refer to such a function as a \emph{Zubov function}. 
Such an approach, initiated in \cite{zubov1964method}, aims to find a Lyapunov function defined on the DOA, valued on $[0, \infty)$, diverging as the state approaches the DOA's boundary, and satisfying a Lyapunov-like inequality. 
From a computational viewpoint, it is more convenient to find a transformed Zubov function valued on $[0, 1]$, so that the $1$-sublevel set is the DOA \cite{liu2025physics, harting2026locally}, or so that the interior of its support set ($0$-superlevel set) is the DOA. 
\begin{lemma}[Zubov's theorem]
    Let $\mbb{O} \subset \mR^d$ be an open set that contains the origin. Then $\mbb{O}$ coincides with the DOA $\mbb{A}$ if and only if there exist functions $\zeta\in C(\mbb O, \mR)$ and $\psi\in C(\mbb O, \mR)$ satisfying the following conditions:
    \begin{enumerate}[label=(\roman*)]
        \item $\zeta(0)=1$ and for all $x\in \mbb{O}\backslash \{0\}$, $\zeta(x)\in (0, 1)$; 
        \item $\psi(0)=0$ and for all $x\in \mbb{O}\backslash \{0\}$, $\psi(x)>0$;
        \item For all $c_3>0$ that is sufficiently small, there exist corresponding $c_1, c_2>0$ such that $1-\zeta(x)>c_1$ and $\psi(x)>c_2$ whenever $|x|\geq c_3$; 
        \item $\zeta(x)\to 0$ as $x\to \tilde x\in \partial \mbb O$, and $\zeta(x)\to 0$ as $|x|\to\infty$ if $\mbb{O}$ is unbounded; 
        \item $\mr{d}^+{\zeta(x)}/\mr{d}t = \psi(x)\zeta(x)$ holds at all $x\in \mbb{O}$. Here the right-hand time derivative is defined as $\lim_{\Delta t\downarrow 0} [\zeta(x(t+\Delta t)) - \zeta(x(t))]/\Delta t$.  
    \end{enumerate}
\end{lemma}

\par The above Zubov's theorem is the statement in \cite{liu2025physics}, rewritten for $\zeta = 1-W$, where $W$ is the Zubov function used there (with $W(0)=0$, $W\to 1$ at $\partial\mbb{O}$ and $\nabla W\cdot f = -\psi(1-W)$). Under this substitution, conditions (iv) and (v) of \cite{liu2025physics} become conditions (iv) and (v) above. We use $\zeta$ rather than $W$ because $\zeta$ is the invariant element of the Zubov--Koopman operator introduced below. Determining a Zubov function $\zeta$ typically takes advantage of a relation of $\zeta$ to the following Lyapunov function, defined as the accumulation of a cost function $\omega \in C(\mR^d,\mR_+)$: 
\begin{equation}\label{eq:Lyapunov}
    v(x) = \int_0^\infty \omega\bra{S_f^t(x)} \xD{t}. 
\end{equation}
To ensure that $v$ is properly defined (finite-valued) on the DOA, we need some regularity conditions regarding the local behavior of this cost integral. 
\begin{assumption}[Regularity of the cost function]
\label{assum:omega}
    The cost function $\omega\in C(\mR^d, \mR_+)$ has the following properties:
    \begin{enumerate}[label=(\roman*)]
        \item $\omega(0)=0$, and $\forall \delta>0$, $\exists c>0$, such that $\omega(x)>c$ whenever $|x|>\delta$;
        \item there exists a $\rho>0$, such that $v(x)<\infty$ ($\forall |x|<\rho$); 
        \item $\forall \epsilon>0$, $\exists \delta>0$, such that $v(x)<\epsilon$ whenever $|x|<\delta$. 
    \end{enumerate}
\end{assumption}
The following proposition was proved in \cite{liu2025physics}. 
\begin{lemma}[Maximal Lyapunov function]
\label{lem:Lyapunov}
    Assume that $f$ is locally Lipschitz on $\mbb{X}$ and $\omega$ satisfies Assumption \ref{assum:omega}. 
    Then $v(0)=0$, $v(x)>0$ ($\forall x\neq 0$), $v(x) < \infty$ ($\forall x\in \mbb A$), $v(x)\rightarrow \infty$ ($x\to \tilde x\in \partial\mbb A$), $v$ is continuous on $\mbb A$, and furthermore $\mr{d}^+v(x)/\mr{d}t = -\omega(x)$. 
\end{lemma}
By transforming the maximal Lyapunov function by an exponential transform, $\zeta(x) = \mr{e}^{-v(x)}$, we obtain a desired candidate for Zubov function.
\begin{lemma}[Zubov function and Zubov PDE]
\label{lem:Zubov}
    Let $\omega$ satisfy Assumption \ref{assum:omega}. Suppose moreover that $f\in \Bra{C^1(\mbb X, \mR)}^d$ and that the origin is an exponentially stable equilibrium point. Then the function $\zeta = \mr{e}^{-v}$, where $v$ is defined in \eqref{eq:Lyapunov}, is a locally Lipschitz function that is the unique viscosity solution to
    \begin{equation*}
        \xPP{\zeta}{x}(x)f(x) = \omega(x)\zeta(x), 
    \end{equation*}
    such that $\zeta(0) = 1$ and $\zeta(x)=0$ on $\partial \mbb{A}$. 
    Furthermore if $\omega\in C^1(\mR^d)$, then $\zeta\in C^1(\mR^d\backslash \partial\mbb{A})$. 
\end{lemma}

\par Now, in view of the relation $\zeta(x) = \mr{e}^{-v(x)}$, we obtain
\begin{equation*}
    (\zeta\circ S_f^t)(x) = \zeta(S_f^t(x)) = \exp\bra{-\int_0^t \omega(S_f^\tau(x)) \mr{d}\tau} \zeta(x). 
\end{equation*}
It is therefore reasonable to define the following operator semigroup, for which $\zeta = \mr{e}^{-v}$ is an invariant element. 
\begin{definition}[Zubov--Koopman semigroup]
\label{def:Zubov-Koopman}
    The operator semigroup $\{Z_{\Delta t}\}_{\Delta t\geq 0}$ given by 
    \begin{equation*}
        (Z_{\Delta t}g)(x) = \exp\bra{-\int_0^{\Delta t} \omega(S_f^\tau(x)) \mr{d}\tau } g(S_f^{\Delta t}(x))
    \end{equation*}
    is referred to as the Zubov--Koopman operator semigroup. 
    The semigroup property is verified as below: for all $s, t\geq 0$, all $g: \mbb{X}\to\mR$, and all $x\in \mbb{X}$, we have
    \begin{equation*}
    \begin{aligned}
        (Z_sZ_tg)(x) &= (Z_s(Z_tg))(x) = \exp\bra{-\int_0^s \omega(S_f^\tau(x)) \xD{\tau}} (Z_tg)(S_f^s(x)) \\
        &= \exp\bra{-\int_0^s \omega(S_f^\tau(x)) \xD{\tau}} \exp\bra{-\int_0^t \omega(S_f^\tau(S_f^s(x))) \xD{\tau}} g(S_f^t(S_f^s(x))) \\
        &= \exp\bra{-\int_0^s \omega(S_f^\tau(x)) \xD{\tau}} \exp\bra{-\int_s^{s+t} \omega(S_f^\tau(x)) \xD{\tau}} g(S_f^{s+t}(x)) \\
        &= \exp\bra{-\int_0^{s+t} \omega(S_f^\tau(x)) \xD{\tau}} g(S_f^{s+t}(x)) = (Z_{s+t}g)(x). 
    \end{aligned}
    \end{equation*}
    This is to say that $Z_sZ_t = Z_{s+t}$ for all $s, t\geq0$. 
\end{definition}

\begin{remark}[Relation to Koopman semigroup]
    With $w_{\Delta t}(x) = \exp\bra{-\int_0^{\Delta t} \omega(S_f^\tau(x)) \mr{d}\tau}$ and the multiplication operator $M_{\mu}: g\mapsto \mu g$, the Zubov--Koopman semigroup is related to the Koopman semigroup by $Z_{\Delta t} = M_{w_{\Delta t}} K_{\Delta t}$.
\end{remark}

It is clear that $Z_{\Delta t}\zeta = \zeta$ for all $\Delta t\geq 0$. However, to compute $\zeta$ from a learned approximation of $Z_{\Delta t}$, one must (i) specify a function space on which $Z_{\Delta t}$ acts and the invariant element is unique, and (ii) show that the computation of the invariant element is stable under perturbations of $Z_{\Delta t}$. These are the subjects of the next two sections.

\section{Zubov--Koopman Operator on the Space of Sectorially Bounded Continuous Functions}\label{sec:operator}
According to Lemma \ref{lem:Zubov}, $\zeta(x) = \exp\bra{-\int_0^\infty \omega(S_f^t(x))\xD{t}} = \lim_{t\rightarrow\infty} w_t(x)$, where we write
\begin{equation}\label{eq:wt}
    v_t(x) = \int_0^t \omega(S_f^\tau(x))\xD{\tau}, \quad w_t(x) = \mr{e}^{-v_t(x)}, \quad \eta_t(x) = 1 - w_t(x).
\end{equation}
Since $w_t = Z_t\mds{1}$, the Zubov function is the pointwise limit of $Z_t\mds{1}$ as $t\to\infty$. For this limit to be computable from a learned operator, three questions must be answered: (i) on which space of functions does the iteration $Z_{\Delta t}^k$ converge, and at which rate; (ii) is $\zeta$ the \emph{unique} invariant element of $Z_{\Delta t}$ on that space (augmented with constants); and (iii) how does a perturbation of $Z_{\Delta t}$, caused by learning from finite data, propagate to the estimate of $\zeta$. This section answers (i) and (ii); Section \ref{sec:learning} answers (iii).

\par The key observation is that every function to which $Z_{\Delta t}$ will be applied (namely $\eta_{\Delta t}$ and $\mds{1}-\zeta$) vanishes at the origin at least linearly. We therefore work on the following Banach space of \emph{sectorially bounded} functions: 
\begin{equation}\label{eq:Cw}
    C_{\mr{sect}}(\mbb{X}) = \BRA{ g\in C(\mbb{X}): g(0)=0, \ |g|_{\mr{sect}} := \sup_{x\in \mbb{X}\backslash\{0\}} \frac{|g(x)|}{|x|} < \infty }.
\end{equation}
Since $\mbb{X}$ is bounded, convergence in $|\cdot|_{\mr{sect}}$ implies uniform convergence, from which the completeness of $C_{\mr{sect}}(\mbb{X})$ follows. Moreover, an error bound in $|\cdot|_{\mr{sect}}$ is exactly a \emph{sectorial} bound, $|g(x)|\leq |g|_{\mr{sect}}|x|$ for all $x\in\mbb{X}$.

\subsection{Exponential Stability of the Zubov--Koopman Semigroup}
The following assumption sets a condition on how large the cost must be, in order that the Zubov--Koopman operator is well-defined and has a nice property about spectral radius. 
\begin{assumption}[Sufficient cost]
\label{assum:large-cost}
	Let $\ell := \sup_{x\in \mbb{X}\backslash\{0\}} x^\top f(x)/|x|^2$ (which satisfies $0<\ell<\infty$ under Assumption \ref{assum:X}). There exist a closed forward-invariant neighborhood $\mbb{O}_1\subseteq \mbb{A}$ of the origin and constants $q\geq 1$, $\beta>0$ such that $|S_f^t(x)|\leq q\mr{e}^{-\beta t}|x|$ for all $x\in\mbb{O}_1$ and $t\geq 0$, and a constant $c>\ell$ such that $\omega(x)\geq c$ for all $x\in \mbb{X}\backslash \mbb{O}_1$. Moreover, $\omega(x)\leq c_\omega |x|$ on $\mbb{X}$ for some $c_\omega>0$.
\end{assumption}

\begin{remark}[On Assumption \ref{assum:large-cost}]
If the Jacobian $J=\mr{D}f(0)$ is Hurwitz, let $P\succ 0$ solve $J^\top P + PJ = -I$. For $r>0$ small enough, $\mbb{O}_1 = \{x: x^\top Px\leq r\}$ is forward invariant, and $|S_f^t(x)|\leq q\mr{e}^{-\beta t}|x|$ on it with $q = \sqrt{\lambda_{\max}(P)/\lambda_{\min}(P)}$. For the commonly used cost $\omega(x) = \alpha|x|^2$, we have $\omega\geq c := \alpha r/\lambda_{\max}(P)$ outside $\mbb{O}_1$, so that $c>\ell$ holds whenever $\alpha$ is large enough. 
The quantity $\ell$ is a one-sided growth rate, which is usually much smaller than the Lipschitz constant of $f$. The condition $c>\ell$ is sufficient but conservative; it is used only to guarantee that trajectories which stay away from $\mbb{O}_1$ for a long time accumulate enough cost to dominate their growth. 
\end{remark}

\begin{lemma}[Spectral radius on $C_{\mr{sect}}$]
\label{lem:Cw.stability}
Under Assumptions \ref{assum:X}, \ref{assum:omega} and \ref{assum:large-cost}, for every $t\geq 0$, $Z_t$ maps $C_{\mr{sect}}(\mbb{X})$ into itself and
\begin{equation*}
    |Z_t g|_{\mr{sect}} \leq q \mr{e}^{-\gamma t}|g|_{\mr{sect}}, \quad \forall g\in C_{\mr{sect}}(\mbb{X}), \quad \gamma := \min\{\beta, \, c-\ell\} > 0.
\end{equation*}
Consequently, the spectral radius of $Z_{\Delta t}$ on $C_{\mr{sect}}(\mbb{X})$ satisfies $r(Z_{\Delta t})\leq \mr{e}^{-\gamma\Delta t}<1$ for every $\Delta t>0$.
\end{lemma}

\subsection{The Zubov Function as the Unique Invariant Element}
We now augment $C_{\mr{sect}}(\mbb{X})$ with the constant function $\mds{1}$, which does not belong to $C_{\mr{sect}}(\mbb{X})$ since it does not vanish at the origin:
\begin{equation}\label{eq:H+1}
    \mc{G}_\oplus = \operatorname{span}\{\mds{1}\}\oplus C_{\mr{sect}}(\mbb{X}), \quad \|a\mds{1}+g\|_{\oplus} := |a| + |g|_{\mr{sect}}.
\end{equation}
We need such an augmented space, since the desired Zubov function, whose support is the domain of attraction, should have a nonzero value at the origin. 
Now because $Z_{\Delta t}\mds{1} = w_{\Delta t} = \mds{1}-\eta_{\Delta t}$, the action of $Z_{\Delta t}$ on $\mc{G}_\oplus$ is
\begin{equation}\label{eq:Z.on.G}
    Z_{\Delta t}(a\mds{1} + g) = a\mds{1} + (Z_{\Delta t}g - a\eta_{\Delta t}), \quad \forall a\in\mR, \ g\in C_{\mr{sect}}(\mbb{X}).
\end{equation}

\begin{theorem}[Zubov function as the unique invariant element]
\label{th:invariant}
Suppose that Assumptions \ref{assum:X}, \ref{assum:omega} and \ref{assum:large-cost} hold and let $\Delta t>0$. Then $\eta_{\Delta t}\in C_{\mr{sect}}(\mbb{X})$, and any invariant element of $Z_{\Delta t}$ on $\mc{G}_\oplus$ is a scalar multiple of
\begin{equation}\label{eq:Zubov.explicit}
    \zeta = \mds{1} - (I-Z_{\Delta t})^{-1} \eta_{\Delta t} = \mds{1} - \sum_{k=0}^\infty Z_{\Delta t}^k \eta_{\Delta t},
\end{equation}
where the series converges in $C_{\mr{sect}}(\mbb{X})$. In this case, $\zeta = \mr{e}^{-v}$; in particular, $\mbb{A}=\{x\in\mbb{X}:\zeta(x)>0\}$. Moreover, $Z_{\Delta t}^{\bar{k}}\mds{1}\to\zeta$ in $\mc{G}_\oplus$ as $\bar{k}\to\infty$, with
\begin{equation*}
    \|Z_{\Delta t}^{\bar{k}}\mds{1}-\zeta\|_\oplus = |Z_{\Delta t}^{\bar{k}}(\mds{1}-\zeta)|_{\mr{sect}} \leq q\mr{e}^{-\gamma \bar{k}\Delta t}|\mds{1}-\zeta|_{\mr{sect}}.
\end{equation*}
\end{theorem}
Theorem \ref{th:invariant} makes the earlier formal relation ``$\zeta = Z_\infty\mds{1}$'' precise: $Z_\infty$ need not exist as an operator, but $Z_{\Delta t}^{\bar{k}}\mds{1}$ converges to $\zeta$ in the norm of $\mc{G}_\oplus$, exponentially fast.

\subsection{Relation to the Linear--Sobolev Space}
The space $C_{\mr{sect}}(\mbb{X})$ is suitable for analysis but not for estimating the Zubov function from data. Learning is performed on the RKHS $\mc{H}:=\mc{H}_{\rg\kappa}(\mbb{X})\equiv\rg H^s(\mbb{X})$, which is continuously embedded in $C_{\mr{sect}}(\mbb{X})$: by the reproducing property and $\rg\kappa(x,x) = |x|^2\rho(0)$,
\begin{equation}\label{eq:embedding}
    |g(x)| = |\ip{g}{\rg\phi(x)}| \leq \|g\|_{\mc{H}}\|\rg\phi(x)\|_{\mc{H}} = \sqrt{\rho(0)}\,|x|\,\|g\|_{\mc{H}}, \quad\text{i.e.,}\quad |g|_{\mr{sect}}\leq\sqrt{\rho(0)}\|g\|_{\mc{H}}.
\end{equation}
The learning analysis requires only that $Z_{\Delta t}$ acts boundedly on $\mc{H}$. 

\begin{assumption}[Boundedness on the RKHS]
\label{assum:bounded}
	Let $s>d/2$, $\eta_{\Delta t}\in\mc{H}$, and $Z_{\Delta t}$ is a bounded operator on $\mc{H}$. Denote $b := \|Z_{\Delta t}\|_{\mc{H}\to\mc{H}}$.
\end{assumption}
\begin{remark}[Sufficient conditions for boundedness on the RKHS] \label{rem:B}
	Let $\mbb{X}$ be star-shaped with respect to the origin, $f\in[C^{\lceil s\rceil+1}(\mbb{X})]^d$ and $\omega\in C^{\lceil s\rceil+1}(\mbb{X})$ with $\omega(0)=0$ and $\nabla\omega(0)=0$. 
	Then $S_f^{\Delta t}(x)=\Sigma(x)x$ and $v_{\Delta t}(x) = \sum_j x_j\iota_j(x)$ with $C^{\lceil s\rceil}$ matrix- and vector-valued $\Sigma$ and $\iota$. Hence, for any $g=\sum_k e_kg_k$,
\begin{equation*}
    Z_{\Delta t}g = \sum_{j=1}^d e_j \sum_{k=1}^d w_{\Delta t}\,\Sigma_{kj}\,(g_k\circ S_f^{\Delta t}).
\end{equation*}
Since multiplication by $C^{\lceil s\rceil}$ functions and composition with $C^{\lceil s\rceil}$ diffeomorphisms are bounded on $H^s(\mbb{X})$ \cite{kohne2025error}, $Z_{\Delta t}$ is bounded on $\rg H^s(\mbb{X})$; similarly $\eta_{\Delta t}\in\rg H^s(\mbb{X})$.
Note, however, that $b<1$ is not guaranteed  as an operator norm on $\rg H^s$, and that, unlike on $C_{\mr{sect}}(\mbb{X})$, the spectral radius of $Z_{\Delta t}$ on $\rg H^s(\mbb{X})$ may exceed $1$. % even if the origin is globally exponentially stable. 
This is why the convergence analysis is carried out on $C_{\mr{sect}}(\mbb{X})$, while only boundedness is required on $\mc{H}$. 
\end{remark}

\section{Operator Learning and Error Analysis}\label{sec:learning}
Throughout this section, $\Delta t>0$ is fixed, and we write $Z = Z_{\Delta t}$ and $\eta = \eta_{\Delta t}$. The cost function $\omega$ is chosen by the user and is therefore known, while the dynamics $f$ are unknown. The dataset consists of $n$ quadruples
\begin{equation}\label{eq:data}
\begin{aligned}
    &\mc{D} = \{(x_i, \, y_i, \, \eta_i, \, \eta_i^+)\}_{i=1}^n, \quad y_i = S_f^{\Delta t}(x_i), \\
    &\eta_i = 1 - \exp\bra{-\int_0^{\Delta t}\omega(S_f^\tau(x_i))\xD\tau}, \quad \eta_i^+ = \eta(y_i) = 1 - \exp\bra{-\int_{\Delta t}^{2\Delta t}\omega(S_f^\tau(x_i))\xD\tau},
\end{aligned}
\end{equation}
with $\{x_i\}_{i=1}^n\subseteq\mbb{X}$. The accumulated costs $\eta_i$ and $\eta_i^+$ are available whenever the trajectory segment from $x_i$ is recorded on $[0,2\Delta t]$ (e.g., at a finer sampling rate, in which case the integral is evaluated by quadrature). If only the states $x_i$, $y_i$ and $S_f^{2\Delta t}(x_i)$ are recorded, the trapezoidal approximation $\eta_i\approx 1-\exp(-\frac{\Delta t}{2}[\omega(x_i)+\omega(y_i)])$ (and similarly for $\eta_i^+$) introduces an error of order $O(\Delta t^3)$ per sample, which can be added to the error bound below. 
We remark that the estimator below will not require evaluating $\eta$ or simulating the system at a new state; instead, the value of $\eta$ at any query point will be replaced by its kernel regression.

\subsection{Learning of the Zubov--Koopman Operator}
The estimation of $Z$ makes use of the following ``push-forward'' relation of its adjoint, which follows from the reproducing property: for all $g\in\mc{H}$, $\ip{g}{Z^*\rg\phi(x)} = (Zg)(x) = (1-\eta(x))g(S_f^{\Delta t}(x))$.
\begin{proposition}[The ``push-forward'' relation of $Z^*$]
\label{prop:Zubov.match}
    Under Assumption \ref{assum:bounded}, for all $x\in \mbb{X}$, $Z^*\rg\phi(x) = (1-\eta(x)) \rg\phi(S_f^{\Delta t}(x))$.
\end{proposition}

Hence, writing $\rg\psi_i = (1-\eta_i)\rg\phi(y_i)$, the operator is estimated by the regularized least-squares problem
\begin{equation}
    \label{eq:regularized.least.squares}
    \hat{Z}^* \in \arg\min_{Z^*} \enspace \sum_{i=1}^n \norm{Z^*\rg\phi(x_i) - \rg\psi_i}_{\mc{H}}^2 + \lambda \|Z^*\|_{\mr{HS}}^2,
\end{equation}
where $\|\cdot\|_{\mr{HS}}$ is the Hilbert--Schmidt norm and $\lambda>0$. Let $C = \sum_{i=1}^n \rg\phi(x_i)\times\rg\phi(x_i)$ be the (unnormalized) sampling operator, where $(h_1\times h_2)g := \ip{h_2}{g}h_1$. Setting the derivative to zero gives $\hat{Z}^* = \bra{\sum_i \rg\psi_i\times\rg\phi(x_i)}(C+\lambda I)^{-1}$. Since $\rg\psi_i = Z^*\rg\phi(x_i)$, we have $\sum_i \rg\psi_i\times\rg\phi(x_i) = Z^*C$, and therefore
\begin{equation}\label{eq:Zhat}
    \hat{Z} = TZ, \quad T := (C+\lambda I)^{-1}C.
\end{equation}
Here $T$ is exactly the \emph{kernel ridge regression} operator: for any $g\in\mc{H}$, $Tg$ is the minimizer of $\sum_i(\hat{g}(x_i)-g(x_i))^2+\lambda\|\hat{g}\|_{\mc{H}}^2$.The operator $T$ is self-adjoint with $0\preceq T\preceq I$; in particular, $\|T\|_{\mc{H}\to\mc{H}}\leq 1$ and $\|\hat{Z}^k\|_{\mc{H}\to\mc{H}}\leq b^k$ for all $k$.

In view of \eqref{eq:Zubov.explicit}, we define the (truncated) estimate of the Zubov function, with truncation level $\bar{k}\in\mN\cup\{\infty\}$, as
\begin{equation}\label{eq:zeta.hat}
    \hat\zeta_{\bar{k}} = \mds{1} - T\eta - \sum_{k=1}^{\bar{k}-1} \hat{Z}^k\eta, \quad \hat{Z}^k\eta = (TZ)^{k-1}\,T(Z\eta).
\end{equation}
The term $k=0$ of the Neumann series, $\eta$ itself, is replaced by its kernel regression $T\eta$, so that $\hat\zeta_{\bar{k}}$ can be evaluated at any $x$ from the data alone. For $k\geq 1$, the regression is applied to $Z\eta$, whose values at the samples, $(Z\eta)(x_i) = (1-\eta_i)\eta(y_i) = (1-\eta_i)\eta_i^+$, are known exactly from the data \eqref{eq:data}. (Replacing $Z\eta$ by $ZT\eta$ would avoid recording $\eta_i^+$, but the regression error of $\eta$ at the successors $y_i$ is then propagated through the Neumann series, which we found to degrade the accuracy considerably in practice.) The following proposition gives the explicit formula, which is proved in the Appendix.
\begin{proposition}[Explicit expression of estimated Zubov function]\label{prop:Zubov.elementary}
	Let $\ms{K} = [\rg\kappa(x_i,x_j)]_{i,j=1}^n$, $\ms{K}_y = [\rg\kappa(y_i,x_j)]_{i,j=1}^n$, $\ms{W} = \operatorname{diag}(1-\eta_1,\ldots,1-\eta_n)$, $\bm{\eta} = [\eta_i]_{i=1}^n$, $\bm{\eta}^+ = [\eta_i^+]_{i=1}^n$, $\ms{k}(x) = [\rg\kappa(x_i,x)]_{i=1}^n$, $\ms{G} = \ms{K}+\lambda\ms{I}$, and $\ms{A} = \ms{G}^{-1}\ms{W}\ms{K}_y$. Then
\begin{equation}\label{eq:Zubov.elementary}
    \hat\zeta_{\bar{k}}(x) = 1 - \ms{k}(x)^\top \Bra{\ms{G}^{-1}\bm{\eta} + \sum_{k=1}^{\bar{k}-1} \ms{A}^{k-1}\ms{G}^{-1}\ms{W}\bm{\eta}^+}.
\end{equation}
If the spectral radius of $\ms{A}$ is less than $1$ (which holds, e.g., when $b<1$), then $\bar{k}=\infty$ is allowed and
\begin{equation}\label{eq:Zubov.elementary.inf}
    \hat\zeta_\infty(x) = 1 - \ms{k}(x)^\top \Bra{\ms{G}^{-1}\bm{\eta} + (\ms{G}-\ms{W}\ms{K}_y)^{-1}\ms{W}\bm{\eta}^+}.
\end{equation}
\end{proposition}

Algorithm \ref{alg:DOA} provides a complete procedure for DOA estimation; the DOA estimated is specified by a threshold value $\vartheta$ on the Zubov function, namely $\hat{\mbb{A}}_\vartheta = \{x\in\mbb{X}: \hat\zeta_{\bar{k}}(x)\geq\vartheta\}$. 
The computational cost of the algorithm is dominated by the factorization of two $n\times n$ matrices and $\bar{k}$ matrix--vector products in the training step, and by $O(n)$ kernel evaluations for each query point.

\begin{algorithm2e}[!htb]
\caption{Kernel Zubov--Koopman estimation of the DOA}\label{alg:DOA}
\KwIn{data $\{(x_i,y_i,\eta_i,\eta_i^+)\}_{i=1}^n$ as in \eqref{eq:data}; radial function $\rho$ and bandwidth $\sigma$; regularization $\lambda>0$; truncation level $\bar{k}$; threshold $\vartheta\in(0,1)$}
\KwOut{Zubov function estimate $\hat\zeta_{\bar{k}}$ and DOA estimate $\hat{\mbb{A}}_\vartheta$}
Form $\ms{K}$, $\ms{K}_y$, $\ms{W}$ with $\rg\kappa(x,x') = (x^\top x')\rho(|x-x'|/\sigma)$, and $\ms{G}=\ms{K}+\lambda\ms{I}$\;
Solve $\ms{G}\,\ms{c}_0=\bm{\eta}$ and $\ms{G}\,\ms{c}_1=\ms{W}\bm{\eta}^+$; set $\ms{A} = \ms{G}^{-1}\ms{W}\ms{K}_y$\;
\For{$k=2,\ldots,\bar{k}-1$}{$\ms{c}_k \leftarrow \ms{A}\,\ms{c}_{k-1}$\;}
Set $\ms{c}=\sum_{k=0}^{\bar{k}-1}\ms{c}_k$ (or $\ms{c} = \ms{c}_0 + (\ms{G}-\ms{W}\ms{K}_y)^{-1}\ms{W}\bm{\eta}^+$ if $\bar{k}=\infty$)\;
\Return{$\hat\zeta_{\bar{k}}(x) = 1-\ms{k}(x)^\top\ms{c}$ and $\hat{\mbb{A}}_\vartheta = \{x\in\mbb{X}:\hat\zeta_{\bar{k}}(x)\geq\vartheta\}$}\;
\end{algorithm2e}

\subsection{Error Analysis}
The error analysis combines the stability of $Z$ on $C_{\mr{sect}}(\mbb{X})$ (Lemma \ref{lem:Cw.stability}), the uniform bound $\|\hat{Z}^k\|_{\mc{H}\to\mc{H}}\leq b^k$, and the accuracy of the regression operator $T$ in the sectorial norm,
\begin{equation}\label{eq:eps.n}
    \varepsilon_n := \|T - I\|_{\mc{H}\to C_{\mr{sect}}(\mbb{X})} = \sup_{\|g\|_{\mc{H}}\leq 1}\, \sup_{x\in\mbb{X}\backslash\{0\}} \frac{|(Tg)(x)-g(x)|}{|x|}.
\end{equation}
The quantity $\varepsilon_n$ is controlled by the fill distance via a sampling inequality.

\begin{definition}[Fill distance]
    The fill distance of $\{x_i\}_{i=1}^n$ in $\mbb{X}$ is 
    $$\ms{h} = \sup_{x\in \mbb{X}} \min_{i=1,\cdots,n} |x-x_i|.$$
\end{definition}
\begin{definition}[Interior cone condition]
    A cone in $\mR^d$ refers to a set of the form
    $\mbb K(x, \xi, \theta, r) = \{x+\lambda y: |y| = 1, y^\top \xi \geq \cos\theta, \lambda \in [0, r]\}$,
    where the vertex $x\in \mR^d$, direction $\xi\in \mR^d$ a unit vector, angle $\theta\in (0, \pi/2)$ and radius $r>0$.
    The set $\mbb X\subseteq \mR^d$ is said to satisfy an interior cone condition if $\exists 0<\theta<\pi/2$ and $r>0$, such that $\forall x\in \mbb X$, $\exists \xi \in \mR^d$ with $\mbb K(x, \xi, \theta, r) \subseteq \mbb X$.
\end{definition}
\begin{remark}[Sufficient condition for interior cone]
    As proven in \cite[Lemma 2.1]{wendland2005approximate}, if $\mbb{X}$ is star-shaped with respect to a ball $\mbb{B}(0, r) = \{x\in \mR^d : |x|<r\}$ and contained in a larger ball $\mbb{B}(0, r')$, then $\mbb{X}$ satisfies the interior cone condition with $r$ and $\theta=2\arcsin(r/2r')$.
\end{remark}

\begin{lemma}[Sectorial bound for kernel ridge regression]
\label{lem:sectorial}
	Suppose that $\mbb{X}$ is bounded, has a Lipschitz boundary, satisfies an interior cone condition, and is star-shaped with respect to the origin, and that $s>1+d/2$. Then there exist constants $c_{\mbb{X}}, \ms{h}_0>0$ depending only on $\mbb{X}$, $s$ and $\rho$, such that whenever $\ms{h}\leq\ms{h}_0$,
	\begin{equation*}
	    \varepsilon_n \leq c_{\mbb{X}}\bra{\ms{h}^{s-1-d/2} + \ms{h}^{-1}\sqrt{\lambda}}.
	\end{equation*}
\end{lemma}

Lemma \ref{lem:sectorial} costs one order of smoothness compared with uniform (non-sectorial) bounds. This is unavoidable in general: a uniform bound on $|Tg-g|$ alone does not control $|Tg(x)-g(x)|/|x|$ near the origin. 
The proof (in the Appendix) therefore should rely on bounding the \emph{gradient} of the regression error.

\begin{theorem}[Sectorial error of the Zubov function estimate]\label{th:uniform}
Suppose that Assumptions \ref{assum:X}--\ref{assum:bounded} hold. Denote $\varsigma_{\bar{k}}(b) := \sum_{m=0}^{\bar{k}-1}b^m$ and $c_\eta := \|\eta\|_{\mc{H}}\bra{1 + q\max\{1,b\}/(1-\mr{e}^{-\gamma\Delta t})}$. Then,
\begin{equation}\label{eq:error.bound}
    |\hat\zeta_{\bar{k}}(x)-\zeta(x)| \leq \Bra{ q\mr{e}^{-\gamma \bar{k}\Delta t}\,|\mds{1}-\zeta|_{\mr{sect}} + c_\eta\,\varsigma_{\bar{k}}(b)\,\varepsilon_n }|x|, \quad \forall x\in\mbb{X}, \quad\forall \bar{k}\in\mN.
\end{equation}
In particular:
\begin{enumerate}[label=(\roman*)]
    \item if $b<1$, then $\bar{k}=\infty$ is allowed and $|\hat\zeta_\infty(x)-\zeta(x)|\leq c_\eta(1-b)^{-1}\varepsilon_n|x|$;
    \item if $b\geq 1$, choosing $\bar{k} = \lceil \log(1/\varepsilon_n)/(\gamma\Delta t + \log b)\rceil$ gives $|\hat\zeta_{\bar{k}}(x)-\zeta(x)|\lesssim \bar{k}\,\varepsilon_n^{\theta}|x|$ with $\theta = \gamma\Delta t/(\gamma\Delta t + \log b)\in(0,1]$.
\end{enumerate}
\end{theorem}

\begin{remark}[Scaling with the sample size and smoothness]
	If the sample points are placed on a quasi-uniform grid, then $\ms{h}\sim n^{-1/d}$; if they are drawn uniformly at random, then $\ms{h}\lesssim(n/\log n)^{-1/d}$ with high probability. 
	With $\lambda\lesssim\ms{h}^{2(s-d/2)}$, Lemma \ref{lem:sectorial} gives $\varepsilon_n\lesssim n^{-(s-1-d/2)/d}$ (up to logarithmic factors), and Theorem \ref{th:uniform} turns this into a sectorial error of the same order when $b<1$, or of order $\varepsilon_n^\theta$ otherwise. 
	Higher smoothness of $f$ and $\omega$ (larger admissible $s$) therefore yields faster convergence, while larger dimension $d$ slows it down.
\end{remark}

\begin{remark}[Choice of hyperparameters]
\label{rem:hyper}
	The method has four hyperparameters: the bandwidth $\sigma$ of the radial function, the cost $\omega$ (here $\omega=\alpha|x|^2$), the regularization $\lambda$, and the truncation level $\bar{k}$. The theory guides their ranges: $\alpha$ must be large enough for Assumption \ref{assum:large-cost}, but a very large $\alpha$ makes $\zeta$ decay steeply and increases $\|\eta\|_{\mc{H}}$; $\lambda$ should scale like $\ms{h}^{2(s-d/2)}$; and $\bar{k}$ is chosen according to Theorem \ref{th:uniform}. The sampling interval $\Delta t$ also matters: a larger $\Delta t$ reduces the number of iterations needed to reach the slow time scales of the dynamics and increases the rate exponent $\theta$ in Theorem \ref{th:uniform}, at the price of longer recorded trajectory segments. 
	Empirically, $(\sigma,\alpha,\lambda)$ are selected by minimizing the misclassification rate on a new validation set of data, i.e., via a cross-validation routine.
\end{remark}

\section{Numerical Examples}\label{sec:example}
To demonstrate the usefulness of the proposed approach, Algorithm \ref{alg:DOA} is verified on four systems, including a planar system with a closed-form Zubov function (Example 1), the reversed van der Pol oscillator (Example 2), a single-machine--infinite-bus power system (Example 3), and a four-dimensional two-machine power system (Example 4). 
All codes for reproducing the results are available at \url{https://github.com/WentaoTang-Pack/DOA-Zubov}.

\subsection{Experimental Protocol and Overall Results}
We use the following protocol consistently on all examples.
\begin{itemize}
    \item \emph{Data sampling.} Training and test states are drawn uniformly from $\mbb{X}_0$ as a disk or a hybercube (box). According to Remark \ref{rem:taper}, the vector field should be multiplied by a smooth cutoff $\chi(|x|)$, equal to $1$ for $|x|\leq r_1$ and to $0$ for $|x|\geq r_2$, with $(r_1,r_2)=(1.4,1.5)$ in Example 1 and $(4.5,5.5)$ in Examples 2--4. % The ball of radius $r_2$ is then forward invariant, and $\mbb{X}_0$ is contained in the region where $\chi=1$.
    Each training sample is a trajectory segment of length $2\Delta t$ from $x_i$, from which $y_i = S_f^{\Delta t}(x_i)$, $\eta_i$ and $\eta_i^+$ are obtained. The cost is $\omega(x) = \alpha|x|^2$.
    A state is labeled as belonging to $\mbb{A}$ if $|x|<10^{-2}$ after simulating the (cutoff) dynamics for $100$ time units. (In Example 1 the labels are exact, $\mbb{A}=\mbb{B}(0,1)$.)
    \item \emph{Performance metrics and training--validation--testing splits.} For each random seed, we draw $n$ training states, $2000$ validation states and $5000$ test states. 
    The estimated DOA is assessed with the following metrics: the false negative rate (FNR, fraction of states in $\mbb{A}$ classified as outside), the false positive rate (FPR, fraction outside $\mbb{A}$ classified as inside), and the balanced accuracy $\mr{BA} = 1-(\mr{FNR}+\mr{FPR})/2$. 
    The threshold $\vartheta$ for specifying the estimated DOA $\hat{\mbb{A}}$ is chosen to maximize the balanced accuracy on the validation set.  
    All reported metrics are computed on the test set. 
    \item \emph{Hyperparameter tuning.} The sampling interval $\Delta t$, cost weight $\alpha$ and bandwidth $\sigma$ were selected once, from $\Delta t\in\{0.05,0.2,0.5,1,2\}$ (up to $6$ in Example 4), $\alpha\in\{0.01,\ldots,10\}$ and $\sigma\in\{1,1.5,2,3\}\times a$, where $a$ is the half-width of $\mbb{X}_0$. 
    The Wendland function $\rho_{d,1}$ is used in the linear--radial kernel, the regularization is $\lambda=10^{-6}$, and $\bar{k}=\infty$ is used whenever $r(\ms{A})<1$ (otherwise $\bar{k}=400$). 
    \item \emph{Baseline.} As a dictionary-based baseline we implemented an EDMD-type estimator in the spirit of \cite{meng2025learning}: with the monomials $\psi$ of total degree at most $p$, a matrix $\ms{M}$ is fitted by ridge least squares to $\psi(x_i)^\top\ms{M} \approx (1-\eta_i)\psi(y_i)^\top$, and $\hat\zeta = \psi^\top\ms{M}^N\ms{e}_1$, where $N=H/\Delta t$. Its hyperparameters $(\Delta t,\alpha,p,H)$ were tuned on the same dataset with the same procedure. 
    \item \emph{Computation.} All computations were run in Python 3.10 (NumPy/SciPy) in a Linux virtual machine with $4$ virtual CPUs and $3$~GB of memory, hosted on an Apple-silicon MacBook Pro laptop. 
    Times are averages over seeds and include the simulation that generates the training data (``data''), the training step (``fit''), and the evaluation of $\hat\zeta$ at $1000$ query states (``predict''). % The cost of labeling the validation and test sets is not part of the method and is not included.
\end{itemize}

\begin{table}
\centering
\caption{Hyperparameters for the examples}\label{tab:hyper}
\begin{tabular}{lccccc}
\toprule
 & $\mbb{X}_0$ & $n$ & Kernel: $(\Delta t, \alpha, \sigma)$ & EDMD: $(\Delta t,\alpha,p,H)$ & $|\mbb{A}\cap\mbb{X}_0|/|\mbb{X}_0|$ \\
\midrule
Ex.~1 & $\mbb{B}(0,1.5)$ & $1000$ & $(0.5,\ 3,\ 3)$ & $(0.5,\ 3,\ 6,\ 50)$ & $0.44$ \\
Ex.~2 & $[-3,3]^2$ & $1000$ & $(2,\ 0.3,\ 6)$ & $(2,\ 0.3,\ 10,\ 50)$ & $0.38$ \\
Ex.~3 & $[-3,3]^2$ & $1000$ & $(0.5,\ 0.3,\ 6)$ & $(2,\ 0.03,\ 6,\ 100)$ & $0.28$ \\
Ex.~4 & $[-2,2]^4$ & $2000$ & $(2,\ 0.03,\ 4)$ & $(6,\ 0.03,\ 6,\ 10)$ & $0.48$ \\
\bottomrule
\end{tabular}
\end{table}

\begin{table}
\centering
\caption{Performance over $20$ random seeds (in $\%$) and mean computational times (in seconds).}\label{tab:main}
\begin{tabular}{ll|ccc|cccc}
\toprule
 & Method & FNR & FPR & BA & data & fit & predict \\
\midrule
Ex.~1 & Kernel (Alg.~1) & $1.0 \pm 0.4$ & $0.6 \pm 0.2$ & $99.2 \pm 0.2$ & $0.02$ & $0.22$ & $0.010$ \\
 & EDMD & $0.4 \pm 0.2$ & $0.2 \pm 0.2$ & $99.7 \pm 0.1$ & $0.02$ & $0.00$ & $0.000$ \\
\midrule
Ex.~2 & Kernel (Alg.~1) & $1.0 \pm 0.4$ & $1.3 \pm 0.5$ & $98.9 \pm 0.2$ & $0.05$ & $0.23$ & $0.009$ \\
 & EDMD & $17.9 \pm 16.0$ & $33.0 \pm 30.8$ & $74.6 \pm 20.2$ & $0.05$ & $0.00$ & $0.001$ \\
\midrule
Ex.~3 & Kernel (Alg.~1) & $2.0 \pm 0.9$ & $4.9 \pm 1.4$ & $96.6 \pm 0.8$ & $0.01$ & $0.25$ & $0.009$ \\
 & EDMD & $5.6 \pm 4.9$ & $52.1 \pm 43.1$ & $71.1 \pm 19.3$ & $0.06$ & $0.00$ & $0.000$ \\
\midrule
Ex.~4 & Kernel (Alg.~1) & $4.7 \pm 1.2$ & $6.2 \pm 1.8$ & $94.6 \pm 0.7$ & $0.15$ & $0.39$ & $0.016$ \\
 & EDMD & $11.2 \pm 4.7$ & $23.9 \pm 15.3$ & $82.4 \pm 8.6$ & $0.46$ & $0.03$ & $0.007$ \\
\bottomrule
\end{tabular}
\end{table}

\subsection{A Simple System with a Closed-Form Zubov Function}
Consider the system
\begin{equation*}
    \dot{x}_1 = -x_1(1-x_1^2-x_2^2) -2\pi x_2, \enspace \dot{x}_2 = -x_2(1-x_1^2-x_2^2) + 2\pi x_1.
\end{equation*}
In polar coordinates, the dynamics can be written as $\dot r = -r(1-r^2)$, $\dot\theta =2\pi$. Hence, the DOA is $\mbb{B}(0, 1)$. 
Solving the radial equation gives $|S_f^t(x)|^2 = |x|^2/[(1-|x|^2)\mr{e}^{2t}+|x|^2]$, so that with $\omega(x)=\alpha|x|^2$,
\begin{equation*}
    \zeta(x) = \mr{e}^{-\int_0^\infty \omega(S_f^\tau(x)) \xD{\tau}} = (1-|x|^2)_+^{\alpha/2}.
\end{equation*}

\begin{figure}
    \centering
    \includegraphics[width=\textwidth]{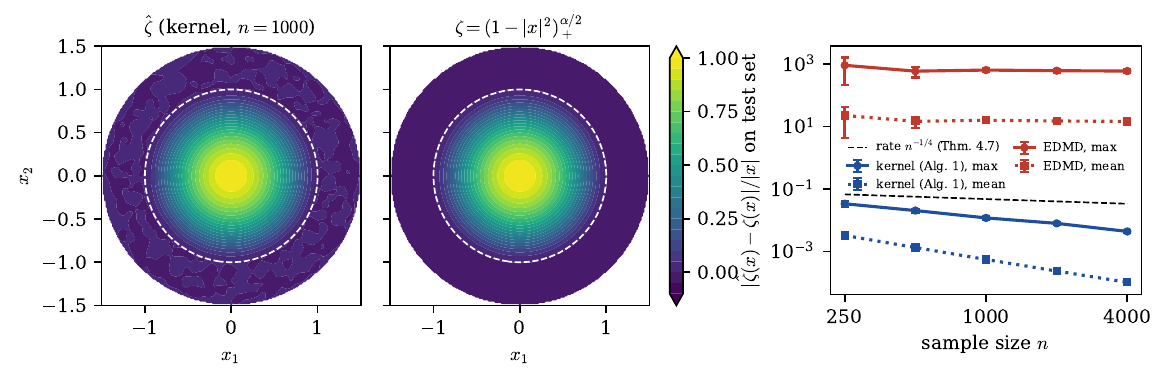}
    \caption{DOA estimation for Example 1. Left: estimated and true Zubov function and the DOA (dashed circles). Right: maximum (solid) and mean (dotted) sectiorial error in $\hat\zeta$.}
    \label{fig:Ex1}
\end{figure}

This example also allows us to check the sectorial error bound of Theorem \ref{th:uniform} directly. Figure \ref{fig:Ex1} compares $\hat\zeta$ and $\zeta$ for $n=1000$ and plots the maximum and mean of $|\hat\zeta(x)-\zeta(x)|/|x|$ over the test set against the sample size. 
The kernel estimate is accurate as a function (test RMSE $1.4\times10^{-3}$, mean sectorial error $5.6\times 10^{-4}$), and both the maximum and the mean sectorial errors decrease with $n$ at least as fast as the predicted rate $n^{-(s-1-d/2)/d} = n^{-1/4}$. Because linear--Sobolev kernel vanishes at the origin, $\hat\zeta(0)=1$ exactly. 
In contrast, the EDMD estimate does not vanish appropriately, its learned matrix has a spectral radius exceeding $1$, and consequently its sectorial error does not decrease with $n$.
The spectral radius of $\ms{A}$ for the kernel method is $0.6065$ in all seeds, which coincides with the value predicted by the Jacobian $r(\mr{e}^{J\Delta t}) = \mr{e}^{-0.5}$.

For DOA classification, both methods perform very well in this example, with balanced accuracies higher than $99\%$. This is expected since $\zeta$ is a function of $|x|^2$ and almost perfectly approximated by low-degree polynomials.

\subsection{Reversed van der Pol Oscillator and Single-Machine--Infinite-Bus Power System}
\begin{figure}
    \centering
    \includegraphics[width=\textwidth]{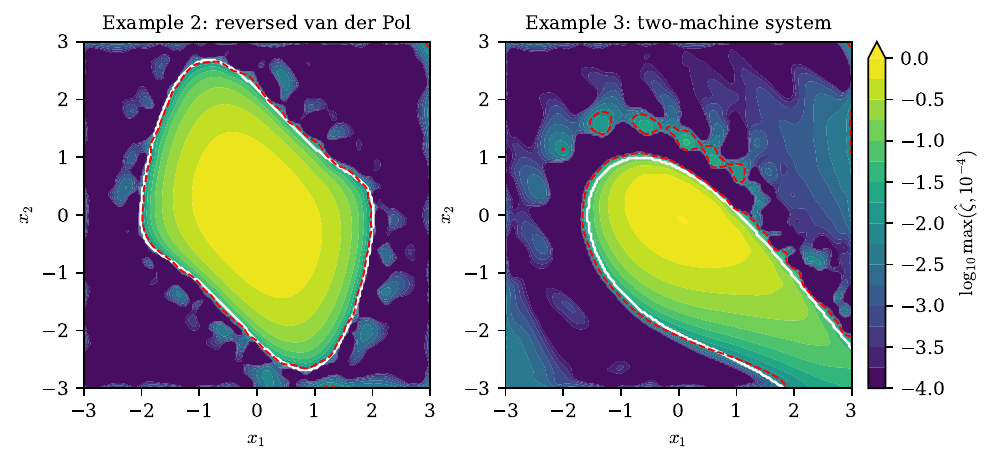}
    \caption{Estimated Zubov functions for Examples 2 and 3. White solid: true DOA boundary from long-horizon simulation; red dashed: boundary of $\hat{\mbb{A}}=\{\hat\zeta\geq\vartheta\}$.}
    \label{fig:Ex23}
\end{figure}

Next we consider a classical polynomial system with a non-trivial DOA, bounded by an unstable limit cycle,
\begin{equation*}
    \dot{x}_1 = x_2, \enspace \dot{x}_2 = -x_1 - (1-x_1^2)x_2,
\end{equation*}
on $\mbb{X}_0 = [-3,3]^2$. Its DOA can be estimated by SOS programming \cite{jones2021converse}, by physics-informed neural networks \cite{liu2025physics}, and by EDMD-based Zubov--Koopman learning \cite{meng2025learning}. 
Figure \ref{fig:Ex23} (left) shows the contour of $\zeta$ (in logarithmic scale), together with the true DOA boundary (white) and the boundary of the estimated set (red, dashed); the two boundaries are nearly indistinguishable. 
Over $20$ seeds, the balanced accuracy of the kernel method is $98.9\pm0.2\%$ (Table \ref{tab:main}), with a training time of $0.23$ seconds. The EDMD baseline attains $74.6\pm20.2\%$ only; its learned matrix has a spectral radius larger than $1$ in all seeds, leading to unstable iterations, and the classification fails ($\mr{BA}<60\%$) in $10$ of the $20$ seeds. 
For reference, \cite{liu2025physics} reported a ``verified volume'' of $95.64\%$ for the PINN approach; this is a different metric (the fraction of the DOA covered by a formally verified sublevel set), obtained with hundreds of seconds of training, and is not directly comparable.

The single-machine--infinite-bus example was used in \cite{meng2025learning} and \cite{liu2025physics} as a benchmark; the experiments in \cite{meng2025learning} showed that SOS programming falls short of obtaining a large DOA estimate. The model is
\begin{equation*}
    \dot x_1 = x_2, \enspace \dot x_2 = -x_2/2 - \bra{\sin (x_1+\theta)-\sin \theta}, \quad \theta = \pi/3,
\end{equation*}
on $\mbb{X}_0=[-3,3]^2$. % Only $28\%$ of $\mbb{X}_0$ belongs to the DOA, whose boundary is formed by stable manifolds of unstable equilibria; states outside it do not converge to the origin. 
As shown in Figure \ref{fig:Ex23} (right), the estimated boundary follows the true one closely; the misclassified states are mainly those in the small spurious ``islands'' above the DOA, which, in principle could be excluded if a connected component detection algorithm is further used. 
The balanced accuracy is $96.6\pm0.8\%$. % In $2$ of the $20$ seeds the spectral radius of $\ms{A}$ slightly exceeds $1$ (at most $1.08$) and the truncated estimator with $\bar k = 400$ is used, without loss of accuracy. 
The EDMD baseline attains $71.1\pm19.3\%$ only, in contrast.

\subsection{Two-Machine Power System (Four States)}
To examine a moderately higher-dimensional case, we consider two generators connected to an infinite bus and to each other:
\begin{equation*}
\begin{aligned}
    \dot x_1 &= x_2, & \dot x_2 &= -d_1x_2 - \bra{\sin(x_1+\theta_1)-\sin\theta_1} - b_{12}\bra{\sin(x_1-x_3+\theta_1-\theta_2)-\sin(\theta_1-\theta_2)}, \\
    \dot x_3 &= x_4, & \dot x_4 &= -d_2x_4 - \bra{\sin(x_3+\theta_2)-\sin\theta_2} + b_{12}\bra{\sin(x_1-x_3+\theta_1-\theta_2)-\sin(\theta_1-\theta_2)},
\end{aligned}
\end{equation*}
where $\theta_1=\pi/3$, $\theta_2 = \pi/4$, $d_1=0.5$, $d_2=0.4$ and $b_{12}=0.5$. 
The origin is exponentially stable (the eigenvalues of $J$ are $-0.219\pm1.236\mr{i}$ and $-0.231\pm0.735\mr{i}$). 
We use $\mbb{X}_0=[-2,2]^4$, of which $48\%$ belongs to the DOA, and $n=2000$ training states. With the dynamics being slow, a larger sampling interval ($\Delta t=2$) was selected, consistent with Theorem \ref{th:uniform}. 

\begin{figure}
    \centering
    \includegraphics[width=0.9\textwidth]{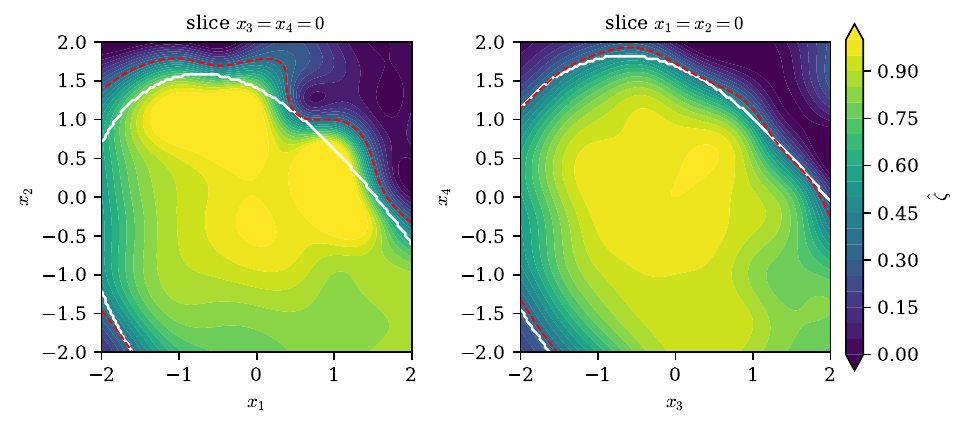}
    \caption{Estimated Zubov function for Example 4. Left: slices in $x_3=x_4=0$. Right: slices in $x_1=x_2=0$. White solid: true DOA boundary; red dashed: boundary of $\hat{\mbb{A}}$.}
    \label{fig:Ex4}
\end{figure}

Figure \ref{fig:Ex4} shows two-dimensional slices of $\hat\zeta$ with the true and estimated DOA boundaries. The balanced accuracy over $20$ seeds is $94.6\pm0.7\%$, with $0.39$ seconds of training, compared with $82.4\pm8.6\%$ for the EDMD baseline (whose learned matrix has a spectral radius of $5.5$ on average, so that its accuracy relies on the short iteration horizon $H=10$). 

\subsection{Effect of the Sample Size}
For the kernel method, the classification error decreases monotonically with $n$ in all four examples, and the variability across seeds is small. 
The training time grows from about $0.01$ seconds ($n=250$) to about $1.6$--$2.2$ seconds ($n=4000$). The EDMD baseline is cheaper computationally, but its accuracy is erratic for small and moderate $n$, due to the instability of the learned EDMD matrix, which fundamentally owes to a lack of spectral theory of EDMD in $C(\mbb{X})$. 
Overall, the results are consistent with the theory: the Zubov function estimate converges as $n$ grows, and the stability of Zubov--Koopman operator on $C_{\mr{sect}}(\mbb{X})$, together with $\|T\|\leq 1$, makes the kernel estimator a robust one. 

\begin{figure}
    \centering
    \includegraphics[width=\textwidth]{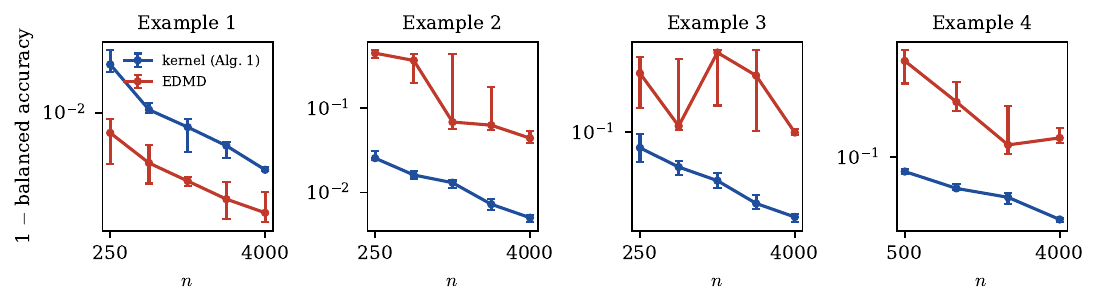}
    \caption{Classification error $1-\mr{BA}$ versus the sample size $n$.}
    \label{fig:sweep}
\end{figure}

\begin{table}
\centering
\caption{Balanced accuracy (\%, mean over $10$ seeds) versus the sample size $n$.}\label{tab:sweep}
\begin{tabular}{llccccc}
\toprule
 & Method & $n=250$ & $500$ & $1000$ & $2000$ & $4000$ \\
\midrule
Ex.~1 & Kernel & $97.6$ & $98.8$ & $99.3$ & $99.4$ & $99.6$ \\
 & EDMD & $99.3$ & $99.6$ & $99.7$ & $99.8$ & $99.8$ \\
\midrule
Ex.~2 & Kernel & $97.3$ & $98.4$ & $98.8$ & $99.3$ & $99.5$ \\
 & EDMD & $62.6$ & $69.2$ & $78.2$ & $86.2$ & $95.2$ \\
\midrule
Ex.~3 & Kernel & $92.2$ & $95.1$ & $96.2$ & $97.4$ & $98.2$ \\
 & EDMD & $68.7$ & $77.4$ & $64.0$ & $70.2$ & $86.1$ \\
\midrule
Ex.~4 & Kernel & -- & $91.8$ & $93.5$ & $94.4$ & $95.8$ \\
 & EDMD & -- & $63.6$ & $73.6$ & $84.8$ & $85.2$ \\
\bottomrule
\end{tabular}
\end{table}

\section{Conclusions}\label{sec:conclusions}
\par In this paper, the Zubov--Koopman operator, whose invariant element is the Zubov function characterizing the DOA, has been shown to be exponentially stable on the space of sectorially bounded continuous functions under a sufficient cost condition. 
Augmenting this space with constants, the Zubov function resides in the resulting space as the unique invariant element of the Zubov--Koopman operator. 
The operator has been learned from snapshot data by kernel ridge regression on the RKHS of a linear--radial product kernel, leading to a closed-form estimate of the Zubov function and an explicit DOA estimation algorithm. 
The error is bounded by a multiple of $|x|$ that decreases with the fill distance of the sample. 

\par As shown in four numerical examples with up to four state dimensions, the proposed method classifies states as inside or outside the DOA with a balanced accuracy of $94.6$--$99.2\%$ using $1000$--$2000$ samples and a training time below $1$ second, and its accuracy improves consistently with the sample size. 
The author postulates that the performance can be even further enhanced by incorporating topological analysis methods. 
For example, as the DOA is known to be a connected open region, the possibility that on-DOA states can be scattered over $\mathbb{X}$ in the midst of off-DOA states should in principle be ruled out. 
Such a numerical phenomenon was seen in the estimated Zubov function profiles in Figure \ref{fig:Ex23} (Example 3), where the scattered islands disconnected from the origin should be further excluded. 

\par The attractors of an autonomous dynamical system, generally, can be much richer than an equilibrium point. The extension of the proposed method to systems with limit cycles, strange attractors, or multiple attractors may be possible by adapting the linear--radial kernel to the corresponding attractors of relevance. 
Yet, from a control-theoretic perspective, a more interesting problem seems to be the estimation of \emph{control-invariant sets}, e.g., null-controllable set, that are maximal under all admissible input actions. 
The problem is also related to that of safety certification using a \emph{control barrier function}, widely used in robotics applications but lacks a general synthesis method.

\acknowledgements{\normalfont Conversations with Alice Harting during the 8\textsuperscript{th} Annual Learning for Dynamics \& Control Conference (L4DC 2026) in Los Angeles made me aware of their interest in solving Zubov problems via neural network approaches. With the belief that kernel methods can offer computationally much more efficient approaches, the author was motivated to complete this work. Claude Code with Opus 5.5 model was used for code revision, hyperparameter tuning, and manuscript editing. The author has reviewed the generated  code and verified its correctness.}

\appendix
\section{Proofs for Propositions and Theorems}
\subsection{Proof of Lemma \ref{lem:Cw.stability}}\label{pf:Cw.stability}
For $x\in\mbb{X}$, let $\tau_{\mr{e}}(x) = \inf\{\tau\geq 0: S_f^\tau(x)\in\mbb{O}_1\}\in[0,\infty]$. By Assumption \ref{assum:large-cost}, $\omega(S_f^\tau(x))\geq c$ for $\tau<\tau_{\mr{e}}(x)$, hence $v_t(x)\geq c\min\{t, \tau_{\mr{e}}(x)\}$.
Let $g\in C_{\mr{sect}}(\mbb{X})$. The function $Z_tg = w_t\cdot(g\circ S_f^t)$ is continuous and $(Z_tg)(0) = g(0) = 0$. For $x\neq 0$, $|(Z_tg)(x)|\leq \mr{e}^{-v_t(x)}|g|_{\mr{sect}}|S_f^t(x)|$. If $t\leq \tau_{\mr{e}}(x)$, then 
\begin{equation*}
    \mr{e}^{-v_t(x)}|S_f^t(x)| \leq \mr{e}^{-ct}\mr{e}^{\ell t}|x| \leq q\mr{e}^{-\gamma t}|x|.
\end{equation*}
If $t>\tau_{\mr{e}}(x)$, then writing $\tau_{\mr{e}}=\tau_{\mr{e}}(x)$ and using the exponential decay on $\mbb{O}_1$,
\begin{equation*}
    \mr{e}^{-v_t(x)}|S_f^t(x)| \leq \mr{e}^{-c\tau_{\mr{e}}}\, q\mr{e}^{-\beta(t-\tau_{\mr{e}})}|S_f^{\tau_{\mr{e}}}(x)| \leq q\mr{e}^{-\beta(t-\tau_{\mr{e}})-(c-\ell)\tau_{\mr{e}}}|x| \leq q\mr{e}^{-\gamma t}|x|.
\end{equation*}
That is, $|Z_tg|_{\mr{sect}}\leq q\mr{e}^{-\gamma t}|g|_{\mr{sect}}$. Since $Z_{\Delta t}^k = Z_{k\Delta t}$ by the semigroup property (Definition \ref{def:Zubov-Koopman}), Gel'fand formula on the Banach space $C_{\mr{sect}}(\mbb{X})$ gives $r(Z_{\Delta t}) = \lim_{k\to\infty}\|Z_{k\Delta t}\|^{1/k}\leq\mr{e}^{-\gamma\Delta t}$. \qed

\subsection{Proof of Theorem \ref{th:invariant}}\label{pf:invariant}
The function $\eta_{\Delta t}=1-\mr{e}^{-v_{\Delta t}}$ is continuous and vanishes at $0$. Using $1-\mr{e}^{-a}\leq a$ for $a\geq 0$, $\omega(x)\leq c_\omega|x|$, we get $0\leq\eta_{\Delta t}(x)\leq v_{\Delta t}(x)\leq c_\omega\Delta t\,\mr{e}^{\max\{\ell,0\}\Delta t}|x|$. Hence $\eta_{\Delta t}\in C_{\mr{sect}}(\mbb{X})$.
By Lemma \ref{lem:Cw.stability}, $r(Z_{\Delta t})<1$ on $C_{\mr{sect}}(\mbb{X})$, so $I-Z_{\Delta t}$ is invertible with $(I - Z_{\Delta t})^{-1}=\sum_{k\geq 0}Z_{\Delta t}^k$. By \eqref{eq:Z.on.G}, $a\mds{1}+g$ is invariant if and only if $(I-Z_{\Delta t})g = -a\eta_{\Delta t}$, i.e., $g = -a(I-Z_{\Delta t})^{-1}\eta_{\Delta t}$. Hence the invariant elements are exactly $a\zeta$, $a\in\mR$.

Since $Z_{\Delta t}^k\mds{1} = w_{k\Delta t}$ and $\eta_{\Delta t} = \mds{1}-Z_{\Delta t}\mds{1}$, the partial sums telescope: $\sum_{k=0}^{\bar{k}-1}Z_{\Delta t}^k\eta_{\Delta t} = \mds{1}-w_{\bar{k}\Delta t}$. As $\bar{k}\to\infty$, the left-hand side converges in $C_{\mr{sect}}(\mbb{X})$, hence pointwise, while $w_{\bar{k}\Delta t}(x) = \mr{e}^{-v_{\bar{k}\Delta t}(x)}\to\mr{e}^{-v(x)}$ monotonically. 
Therefore $\zeta = \mr{e}^{-v}$ on $\mbb{X}$, and $Z_{\Delta t}^{\bar{k}}\mds{1}-\zeta = w_{\bar{k}\Delta t}-\zeta = -\sum_{k\geq \bar{k}}Z_{\Delta t}^k\eta_{\Delta t} = -Z_{\Delta t}^{\bar{k}}(\mds{1}-\zeta)$, which yields the stated rate by Lemma \ref{lem:Cw.stability}. 
Finally, we show that $\mbb{A} = \{\zeta>0\}$. If $x\notin\mbb{A}$, then $\tau_{\mr{e}}(x)=\infty$, $v_t(x)\geq ct\to\infty$ and $\zeta(x)=0$. If $x\in\mbb{A}$, then $\tau_{\mr{e}}(x)<\infty$, and $v(x) \leq v_{\tau_{\mr{e}}(x)}(x) + \int_0^\infty c_\omega q\mr{e}^{-\beta\tau} |S_f^{\tau_{\mr{e}}(x)}(x)| \xD\tau <\infty$, implying that $\zeta(x)>0$. \qed

\subsection{Proof of Proposition \ref{prop:Zubov.elementary}}\label{pf:Zubov.elementary}
For $g = \sum_{j}\ms{c}_j\rg\kappa(x_j,\cdot)$, we have $(Zg)(x_i) = (1-\eta_i)g(y_i) = (\ms{W}\ms{K}_y\ms{c})_i$. The regression $T$ maps any $h\in\mc{H}$ to $\sum_j\ms{c}'_j\rg\kappa(x_j,\cdot)$ with $\ms{c}' = \ms{G}^{-1}[h(x_i)]_{i=1}^n$. 
Hence $\mc{H}_n = \operatorname{span}\{\rg\kappa(x_j,\cdot)\}_{j=1}^n$ is invariant under $\hat Z = TZ$, and in the coefficients of $\mc{H}_n$, $\hat Z$ acts as $\ms{c}\mapsto\ms{A}\ms{c}$. The function $T\eta$ has coefficients $\ms{G}^{-1}\bm{\eta}$, and $T(Z\eta)$ has coefficients $\ms{G}^{-1}\ms{W}\bm{\eta}^+$ since $(Z\eta)(x_i) = (1-\eta_i)\eta_i^+$. As $\hat{Z}^k\eta = (TZ)^{k-1}T(Z\eta)$, \eqref{eq:Zubov.elementary} follows by evaluating at $x$ with the reproducing property. If $r(\ms{A})<1$, $\sum_{k\geq 1}\ms{A}^{k-1} = (\ms{I}-\ms{A})^{-1}$, and $(\ms{I}-\ms{A})^{-1}\ms{G}^{-1} = (\ms{G} - \ms{W}\ms{K}_y)^{-1}$ gives \eqref{eq:Zubov.elementary.inf}. Finally, $\ms{A}$ represents the restriction of $\hat{Z}$ to the invariant subspace $\mc{H}_n$, so $r(\ms{A})\leq r(\hat{Z})\leq\|\hat Z\|_{\mc{H}\to\mc{H}}\leq b$. \qed

\subsection{Proof of Lemma \ref{lem:sectorial}}\label{pf:sectorial}
Fix $g\in\mc{H}$ with $\|g\|_{\mc{H}}\leq 1$ and let $u = Tg-g\in\mc{H}$, so that $u(0)=0$ and $\|u\|_{\mc{H}}\leq\|T\|\|g\|+\|g\|\leq 2$. Moreover, $\mc{H}\equiv\rg H^s(\mbb{X})$ embeds continuously into $H^s(\mbb{X})$: for any representation $u=\sum_ke_ku_k$, $\|u\|_{H^s}\leq\sum_k\|e_ku_k\|_{H^s}\leq c_e\sum_k\|u_k\|_{H^s}\leq c_e\sqrt{d}\,(\sum_k\|u_k\|_{H^s}^2)^{1/2}$, hence $\|u\|_{H^s}\leq c_e\sqrt{d}\|u\|_{\mc{H}}$. 
Comparing the ridge objective at its minimizer $Tg$ with its value at $g$ gives $\sum_i|u(x_i)|^2 + \lambda \|Tg\|_{\mc{H}}^2 \leq \lambda\|g\|_{\mc{H}}^2$, hence $\max_i|u(x_i)|\leq\sqrt{\lambda}$. 
Since $s>1+d/2$, the sampling inequality for functions in $H^s(\mbb{X})$ on domains satisfying an interior cone condition \cite[Theorem 2.6]{wendland2005approximate} gives, for $\ms{h}\leq\ms{h}_0$,
\begin{equation*}
    \|\nabla u\|_{L^\infty(\mbb{X})} \leq c_{\mbb{X}}' \bra{\ms{h}^{s-1-d/2} \|u\|_{H^s(\mbb{X})} + \ms{h}^{-1}\max_i|u(x_i)|}.
\end{equation*}
As $\mbb{X}$ is star-shaped with respect to $0$ and $u(0)=0$, $|u(x)| = |\int_0^1\nabla u(\theta x)^\top x\,\xD\theta|\leq\|\nabla u\|_{L^\infty}|x|$.

\subsection{Proof of Theorem \ref{th:uniform}}\label{pf:uniform}
By \eqref{eq:Zubov.explicit} and \eqref{eq:zeta.hat},
\begin{equation*}
    \hat\zeta_{\bar{k}}-\zeta = \sum_{k\geq \bar{k}}Z^k\eta - \sum_{k=0}^{\bar{k}-1}\delta_k, \quad \delta_0 := (T-I)\eta, \quad \delta_k := \hat{Z}^k\eta - Z^k\eta \ (k\geq 1),
\end{equation*}
where $\sum_{k\geq \bar{k}}Z^k\eta = Z^{\bar{k}}\sum_{k\geq 0}Z^k\eta = Z^{\bar{k}}(\mds{1}-\zeta)$, whose $|\cdot|_{\mr{sect}}$-norm is at most $q\mr{e}^{-\gamma \bar{k}\Delta t}|\mds{1}-\zeta|_{\mr{sect}}$ by Lemma \ref{lem:Cw.stability}.
First, $|\delta_0|_{\mr{sect}}\leq\varepsilon_n\|\eta\|_{\mc{H}}$. For $k\geq 1$, we note that $\hat{Z}^k - Z^k = \sum_{j=0}^{k-1}Z^{j}(\hat{Z}-Z)\hat{Z}^{k-1-j}$ with $\hat{Z}-Z = (T-I)Z$. For every $m$, $Z\hat{Z}^m\eta\in\mc{H}$ with $\|Z\hat{Z}^m\eta\|_{\mc{H}}\leq b^{m+1}\|\eta\|_{\mc{H}}$ because $\|T\|_{\mc{H}\to\mc{H}}\leq 1$; the operator $T-I$ maps it into $C_{\mr{sect}}(\mbb{X})$ with norm at most $\varepsilon_n$; and $Z^{j}$ acts on $C_{\mr{sect}}(\mbb{X})$ with norm at most $q\mr{e}^{-\gamma j\Delta t}$. Hence
\begin{equation*}
    |\delta_k|_{\mr{sect}}\leq\varepsilon_n\|\eta\|_{\mc{H}}\sum_{j=0}^{k-1}q\mr{e}^{-\gamma j\Delta t}b^{k-j}, \quad k\geq 1.
\end{equation*}
Summing over $1\leq k<\bar{k}$, the double sum is bounded by the product of $\sum_{j\geq 0}q\mr{e}^{-\gamma j\Delta t} = q/(1-\mr{e}^{-\gamma\Delta t})$ and $\sum_{m=1}^{\bar{k}-1}b^m\leq\max\{1,b\}\varsigma_{\bar{k}}(b)$, which, together with $\varsigma_{\bar{k}}(b)\geq 1$, gives $\sum_{k<\bar{k}}|\delta_k|_{\mr{sect}}\leq c_\eta\varsigma_{\bar{k}}(b)\varepsilon_n$. Adding the two terms proves \eqref{eq:error.bound}. Case (i) follows from $\varsigma_\infty(b)=(1-b)^{-1}$. For case (ii), $\varsigma_{\bar{k}}(b)\leq \bar{k}b^{\bar{k}}$, and the choice of $\bar{k}$ balances $\mr{e}^{-\gamma \bar{k}\Delta t}$ and $\varepsilon_nb^{\bar{k}}$, both of which are then of order $\varepsilon_n^\theta$. \qed

\end{document}